\documentclass{article}
\usepackage{latexsym, amsmath, amssymb, mathrsfs,
  graphics, color, multicol, epsfig} 
\usepackage{epstopdf}
\def \uqsl2 {$U_q$(sl$_2$) }
\def \cA {{\cal A}}

\newcommand {\be}{\begin{equation}}
\newcommand {\ee} {\end{equation}}
\newcommand {\bea}{\begin{eqnarray}}
\newcommand {\eea} {\end{eqnarray}}
\makeatletter
\@addtoreset{equation}{section}
\makeatother

\def\bea{\begin{eqnarray}}
\def\eea{\end{eqnarray}}
\def\be{\begin{equation}}
\def\ee{\end{equation}}

\def\wgta#1#2#3#4{\hbox{\rlap{\lower.35cm\hbox{$#1$}}
\hskip.2cm\rlap{\raise.25cm\hbox{$#2$}}
\rlap{\vrule width1.3cm height.4pt}
\hskip.55cm\rlap{\lower.6cm\hbox{\vrule width.4pt height1.2cm}}
\hskip.15cm
\rlap{\raise.25cm\hbox{$#3$}}\hskip.25cm\lower.35cm\hbox{$#4$}\hskip.6cm}}

\def\wgtb#1#2#3#4{\hbox{\rlap{\raise.25cm\hbox{$#2$}}
\hskip.2cm\rlap{\lower.35cm\hbox{$#1$}}
\rlap{\vrule width1.3cm height.4pt}
\hskip.55cm\rlap{\lower.6cm\hbox{\vrule width.4pt height1.2cm}}
\hskip.15cm
\rlap{\lower.35cm\hbox{$#4$}}\hskip.25cm\raise.25cm\hbox{$#3$}\hskip.6cm}}

\def\begeqar{\begin{eqnarray}}
\def\endeqar{\end{eqnarray}}
\title{Combinatorial aspects of boundary loop models}
\author{Jesper Lykke Jacobsen${}^{1,2}$ and
        Hubert Saleur${}^{2,3}$ \\[2.0mm]
${}^1$ LPTMS, Universit\'e Paris-Sud, B\^atiment 100, \\
Orsay, 91405, France \\
${}^2$ Service de Physique Th\'eorique, CEA Saclay, \\
Gif Sur Yvette, 91191, France \\
${}^3$ Department of Physics and Astronomy,
University of Southern California, \\
Los Angeles, CA 90089, USA}

\begin{document}

\maketitle

\begin{abstract}

We discuss in this paper combinatorial aspects of boundary loop models, that is models of self-avoiding loops on a strip where loops get different weights depending on whether they touch the left, the right, both or no boundary. These models are described algebraically by a generalization of the Temperley-Lieb algebra, dubbed the two-boundary TL algebra. We give results for the dimensions of TL representations and the corresponding degeneracies in the partition functions. We interpret these results in terms of fusion and in the light of the recently uncovered ${\cal A}_n$ large symmetry present in loop models, paving the way for the analysis of the conformal field theory properties. Finally, we propose conjectures for determinants of Gram matrices in all cases, including the two-boundary one, which has recently been discussed by de Gier and Nichols.

\end{abstract}

%
%

\section{Introduction}

The study of models of self-avoiding loops plays an important role in
two-dimensional statistical physics and makes contact with a variety of
fields: conformal field theory, integrable systems, algebra and representation
theory, combinatorics, and probability theory.

\begin{figure}
  \begin{center}
    \includegraphics[scale=0.4]{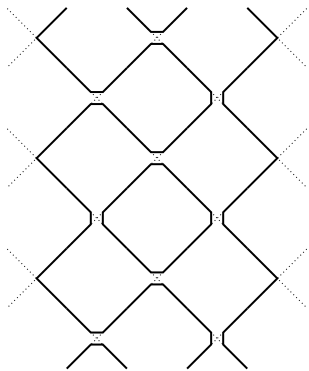}
  \end{center}
  \caption{Configuration of self-avoiding fully-packed loops on an annulus.}
  \label{fig:config}
\end{figure}

A particularly rich and interesting model is obtained by covering all the
edges of a (tilted) square lattice with loops which split in one of two
possible ways at each bulk vertex, and are reflected at boundary vertices. A
possible configuration is shown in Fig.~\ref{fig:config}. We shall focus on
the case where the overall topology is that of an annulus, with reflecting
boundary conditions horizontally, and periodic boundary conditions vertically.
The loops can then have two different homotopies with respect to the periodic
direction: contractible or non-contractible. In general, we may define a
statistical ensemble by giving a weight $n$ to each contractible loop and a
weight $\ell$ to each non-contractible loop. The weight of the configuration
in Fig.~\ref{fig:config} is then $n \ell^2$.

The model just defined is closely related to the $Q$-state Potts model, and
has the algebraic structure of the Temperley-Lieb algebra. It also has a
\uqsl2 quantum group symmetry.

In Ref.~\cite{JS} we have initiated the study of a more general model in which
special weights $n_{\rm l}$ and $\ell_{\rm l}$ are given to loops that touch
at least once the left boundary. This model has the algebraic structure of the
blob algebra (introduced in \cite{MS}), which has also been called the
one-boundary Temperley-Lieb (1BTL) algebra in recent papers. This latter
terminology means that only one of the two boundaries (by convention, the left
one) gives rise to special loop weights. In the same spirit, we shall
sometimes refer to the ordinary Temperley-Lieb case (with no distinguished
boundary) as 0BTL.

In our first paper \cite{JS}, we have mostly elucidated physical features of
the one-boundary case, the most mathematical ones having been studied in
earlier works \cite{MathBlob,NRG,N}. In the limit of an infinitely large
lattice, the model turns out to be conformally invariant, and each choice of
$n_{\rm l}$ gives rise to a distinct conformal boundary condition.

In this second paper, we deal with some combinatorial aspects of the
two-boundary case, in preparation for the study of its conformal field
theoretic aspects. Our wish to single out the combinatorial treatment is that
(most of) the arguments given are completely rigorous, and the results hold
true for a lattice of any finite size. Moreover, they are to a large extent
independent of the underlying lattice structure, and of the particularities of
the model at hand (fully-packing constraint, choice of {\em local} vertex
weights, etc). They underlying algebraic framework is now the two-blob---or
two-boundary Temperley Lieb (2BTL)---algebra \cite{N,Ni,GN}. This algebra
raises fascinating questions, both of a mathematical nature (representation
theory, etc) and of a more physical nature (boundary conformal field theory,
etc).

In full generality, the weight of a loop in the two-boundary case is then
given by the following table:
\begin{equation}
\begin{tabular}{lll|l}
 Contractible & Touches the   & Touches the    & Weight        \\
              & left boundary & right boundary &               \\ \hline
 Yes          & No            & No             & $n$           \\
 Yes          & Yes           & No             & $n_{\rm l}$    \\
 Yes          & No            & Yes            & $n_{\rm r}$    \\
 Yes          & Yes           & Yes            & $n_{\rm b}$    \\
 No           & No            & No             & $\ell   $     \\
 No           & Yes           & No             & $\ell_{\rm l}$ \\
 No           & No            & Yes            & $\ell_{\rm r}$ \\
 No           & Yes           & Yes            & $\ell_{\rm b}$ \\
\end{tabular}
\label{loopweights}
\end{equation}
It is also appropriate to gather here some other notations that we shall
use throughout the paper:
\begin{equation}
\begin{tabular}{ll}
 $N$   & Strip width (number of strands/sites), sometimes written $N=2N_2$ \\
 $L$   & Number of non-contractible lines \\
 $e_i$ & Temperley-Lieb (bulk) generator \\
 $b_{\rm l},b_{\rm r}$ & Blob (boundary) generators \\
 $D$   & Dimension of the commutant (eigenvalue amplitude) \\
 $d$   & Dimension of invariant subspace (transfer matrix dimension) \\
 $U_L(\cos\theta)={\sin(L+1)\theta\over \sin\theta}$
       & $L$th order Chebyshev polynomial of the second kind \\
 ${p \choose q}$
       & Binomial coefficient $\frac{p!}{q!(p-q)!}$ for integer
         $0 \le q \le p$; zero otherwise \\
\end{tabular}
\label{notation}
\end{equation}
Note that, due to the fully-packing constraint, $N$ and $L$ must have the same
parity. In Fig.~\ref{fig:config}, $N=4$ and $L=2$. Throughout we shall assume
$N$ even, except when the contrary is stated explicitly. This restriction is
not essential, and it is imposed mostly in order to write down the simplest
formulae only.

In sections 2--4 we shall introduce the transfer matrix of the
loop model, define the states it acts on, relate it to the underlying
algebras, define and compute the dimensions $d$ of its various sectors, and
derive the corresponding eigenvalue amplitudes $D$ (which are also the
dimensions of the commutant). To gain clarity, we shall do so gradually, by
treating first the ordinary Temperley-Lieb case (0BTL), and then consider next
the one- and two-boundary cases.

In section 5 we give a more algebraic account on the numbers $D$, and in
section 6 we discuss the determinants of various Gram matrices which occur
naturally when studying the representation theory of the boundary
Temperley-Lieb algebras. Finally, section 7 is devoted to our conclusions.

\section{Zero-boundary case}

Although the 0BTL case has long been well understood
\cite{Martin,Jones,BauerSaleur}, we shall review it here since it contains many of
the elements necessary to attack the cases with boundaries. This
approach also has the advantage of fixing a consistent notation and
terminology which should facilitate the reading of sections 3--4.

\subsection{Algebraic structure}

\begin{figure}
  \begin{center}
    \includegraphics[scale=0.4]{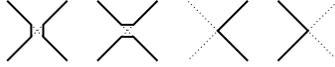}
  \end{center}
  \caption{From left to right: identity $I$ and Temperley-Lieb generator $e_i$
    acting on two strands $i$ and $i+1$; left and right boundary identity
    operator.}
  \label{fig:vertices}
\end{figure}

Consider a system of $N$ strands labeled $i=1,2,\ldots,N$. The lattice is
built up from elementary generators $e_i$, acting on strands $i$ and $i+1$, as
shown in Fig.~\ref{fig:vertices}. More precisely, in the case where all {\em
  local} vertex weights are unity, the transfer matrix reads
\begin{equation}
 T = \left( \prod_{j=1}^{N/2-1} (I + e_{2j}) \right)
     \left( \prod_{j=1}^{N/2} (I + e_{2j-1}) \right)
\end{equation}

The generators $e_i$ satisfy the well-known relations
\begin{eqnarray}
 e_i^2&=&ne_i\nonumber\\
 e_i e_{i \pm 1}e_i &=& e_i\nonumber\\
 \left[e_i,e_j\right]&=&0 \mbox{ for $|i-j|\geq 2$}
 \label{TL0}
\end{eqnarray}
The identity and the $N-1$ generators $e_i$ define the Temperley-Lieb algebra
$TL_{N}(n)$, subject to the above relations. Graphically, the application of
the last two relations allows to deform and diminish the size of a loop, and
when it has reached its minimal possible size it can be taken away and
replaced by the weight $n$ due to the first relation.

\subsection{States and transfer matrix decomposition}
\label{sec:states0}

\begin{figure}
  \begin{center}
    \includegraphics[scale=0.4]{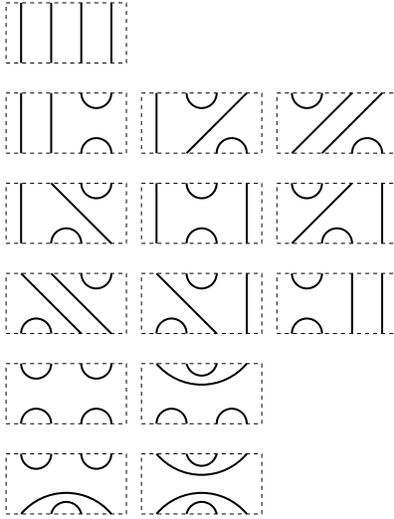}
  \end{center}
  \caption{List of all 0BTL states on $N=4$ strands. Each row corresponds
to a definite sector of the transfer matrix.}
  \label{fig:states0}
\end{figure}

The transfer matrix $T$ acts on states which can be depicted graphically as
non-crossing link patterns within a slab bordered by two horizontal rows, each
of $N$ points. The complete list of states for $N=4$ is shown in
Fig.~\ref{fig:states0}. The bottom (resp.\ top) row of the slab corresponds to
time $t=0$ (resp.\ $t=t_0$); the transfer matrix propagates the states from
$t_0$ to $t_0+1$ and thus acts on the top of the slab only.

\begin{figure}
  \begin{center}
    \includegraphics[scale=0.4]{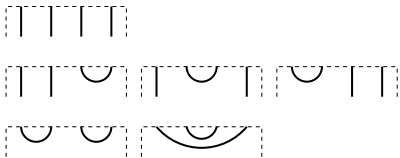}
  \end{center}
  \caption{List of all 0BTL reduced states on $N=4$ strands. Each row
    corresponds to a definite sector of the transfer matrix.}
  \label{fig:red_states0}
\end{figure}

A link joining the top and the bottom of the slab is called a {\em string},
and any other link is called an {\em arc}. We denote by $s$ the number of strings
in a given state. Any state can be turned into a pair of {\em reduced states}
by cutting all its strings and pulling apart the upper and lower parts. For
convenience, a cut string will still be called a string with respect to the
reduced state. The complete list of reduced states for $N=4$ is shown in
Fig.~\ref{fig:red_states0}.

Conversely, a state can be obtained by adjoining two reduced states, gluing
together their strings in a unique fashion. Thus, if we define $d_j$ as the
number of reduced states with $s=2j$ strings, the number of states with $s=2j$
strings is simply $d_j^2$.

The partition function $Z_{N,M}$ on an annulus of width $N$ strands and height
$M$ units of time cannot be immediately expressed in terms of reduced states
only, since these do not contain the information about how many loops
(contractible or non-contractible) are formed when the periodic boundary
condition is imposed. We can however write it in terms of states as
\begin{equation}
 Z_{N,M} = \langle u | T^M | v \rangle \,.
\end{equation}
At time $t_0=0$ the top and the bottom of the slab must be identified.
Therefore, the entries of the right vector $| v \rangle$ are one
whenever the corresponding state contains no arcs, and each of its
links connects a point in the bottom row to the point immediately
above it in the top row; all other entries of $| v \rangle$ are zero.
At time $t_0=M$ the top and the bottom of the slab must be reglued.
Therefore, the left vector $\langle u |$ is obtained by identifying
the top and bottom rows for each state; counting the number of loops
of each type gives the corresponding weight as a monomial in the loop
weights $n$ and $\ell$.

The reduced states can be ordered according to a decreasing number of strings.
The states can be ordered first according to a decreasing number of strings,
and next, for a fixed number of strings, according to its bottom half reduced
state. These orderings are brought out by the rows in
Figs.~\ref{fig:states0}--\ref{fig:red_states0}.

With this ordering of the states, $T$ has a blockwise lower triagonal
structure in the basis of reduced states, since the generator $e_i$ can
annihilate two strings (if their position on the top of the slab are $i$ and
$i+1$) but cannot create any strings.

In the basis of states, $T$ is blockwise lower triagonal with respect to the
number of strings, for the same reason. Each block on the diagonal in this
decomposition corresponds to a definite number of strings. The block
corresponding to $s=2j$ strings is denoted $\tilde{T}_j$. But since $T$ acts
only on the top of the slab, each $\tilde{T}_j = T_j \oplus \ldots \oplus T_j$
is in turn a direct sum of $d_j$ identical blocks $T_j$ which correspond
simply to the action of $T$ on the reduced states with $2j$ strings.

In particular, the eigenvalues of $T$ are the union of the eigenvalues of
$T_j$, where the $T_j$ now act in the much smaller basis of reduced states.
This observation is particularly useful in numerical studies.

\subsection{The dimensions $d_L$ and $D_L$}

We briefly review a combinatorial construction \cite{RJ} which we shall
generalize to the case with boundaries in the following sections.
We take for now  the width of the annulus $N=2N_2$  to be even.
For each transfer matrix block $T_j$ we define the corresponding character
as
\begin{equation}
 K_j = {\rm Tr}\, \left( T_j \right)^M \,,
\end{equation}
where we stress that the trace is over {\em reduced} states. Obviously we
have
\begin{equation}
 K_j = \sum_{i=1}^{d_j} \left( \lambda_i^{(j)} \right)^M \,,
 \label{defK0}
\end{equation}
where $\lambda_i^{(j)}$ are the eigenvalues of $T_j$. The expression of the
partition function in terms of transfer matrix eigenvalues is more involved,
due essentially to the non-local nature of the loops, and reads
\begin{equation}
 Z_{N,M} \equiv \sum_{j=0}^{N_2} Z_j = \sum_{j=0}^{N_2} D_{2j} K_j \,,
 \label{ZDK}
\end{equation}
where $Z_j$ is the annulus partition function constrained to have exactly
$L=2j$ non-contractible loops, and $D_{2j}$ are some eigenvalue amplitudes to be
determined. To be more precise, we decompose $Z_j$ in terms of $K_k$ as
follows
\begin{eqnarray}
 Z_j &=& \sum_{k=j}^{N_2} D(k,j) \ell^{2k} K_k \\
 D_{2j} &=& \sum_{i=0}^j D(j,i) \ell^{2i} \nonumber
\end{eqnarray}
and consider next the inverse decomposition
\begin{equation}
 K_k = \sum_{j=k}^{N_2} E(j,k) \frac{Z_j}{\ell^{2j}} \,.
 \label{inv_decomp_0}
\end{equation}

\begin{figure}
  \begin{center}
    \leavevmode
    \epsfysize=60mm{\epsffile{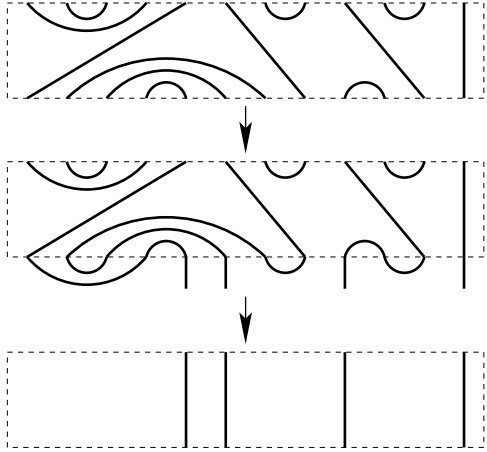}}
  \end{center}
  \protect\caption{Construction of invariant reduced states.
    (a) A configuration contributing to $Z_2$ with $N_2=6$, here
    depicted as a state. (b) Application on the bottom of the reduced
    state corresponding to the top half of (a). (c) After removal of
    the arcs one has simply $2j$ links.}
  \label{compstate}
\end{figure}

The determination of the coefficients $E(j,k)$ can be turned into a
combinatorial counting problem as follows. First, recall that the characters
$K_k$ were defined as {\em traces} over reduced states. We must now
determine how many times each $Z_j$ occurs within a given trace. Consider
therefore some configuration ${\cal C}$ on the annulus that contributes to
$Z_j$. An example with $j=2$ and $N_2=6$ is shown in Fig.~\ref{compstate}a. It
is convenient not to represent the contractible loops within the
configuration, i.e., to depict it as a state. This configuration will
contribute to the trace only over such reduced states ${\cal S}$ that are
left {\em invariant} by the action of the configuration. Therefore, ${\cal S}$
must contain the same arcs as does ${\cal C}$ in its top row (see
Fig.~\ref{compstate}b). It suffices therefore to determine the parts of ${\cal
  S}$ which connect onto the starting points of the $2j$ non contractible
lines (see Fig.~\ref{compstate}c). Since the goal is to determine the
contribution to $K_k$, precisely $2k$ strings and $j-k$ arcs must be used.
In other words, $E(j,k)$ is precisely the number of reduced states on $2j$
strands, and using $2k$ strings.

Now let
\begin{equation}
 E^{(k)}(z) = \sum_{j=0}^\infty E(j,k) z^j
\end{equation}
be the corresponding generating function, where $z$ is a formal parameter
representing the weight of an arc, or of a pair of strings. When $k=0$, a
reduced state with no strings is either empty, or has a leftmost arc which
divides the space into two parts (inside the arc and to its right) each of
which can accomodate an independent arc state. The generating function $f(z)
\equiv E^{(0)}(z)$ therefore satisfies $f(z) = 1 + z f(z)^2$ with regular
solution
\begin{equation}
 f(z) = \frac{1 - \sqrt{1-4z}}{2z} = \sum_{j=0}^\infty 
 \frac{(2j)!}{j!(j+1)!} \, z^j \,.
 \label{gen_f}
\end{equation}
When $k \neq 0$, the strings simply divide the space into $2k+1$ parts each of
which contains an independent arc state. Therefore,
\begin{equation}
 E^{(k)}(z) = z^k f(z)^{2k+1} = \sum_{j=k}^\infty
 \left[ {2j \choose j+k} - {2j \choose j+1+k} \right] z^j
\end{equation}
and in particular we have
\begin{equation}
 d_L = E \left( \frac{N}{2},\frac{L}{2} \right)
     = {N \choose (N+L)/2} - {N \choose 1+(N+L)/2} \,.
 \label{dL0}
\end{equation}
Note that $d_L$ depends on $N$, but we usually will not mention this
explicitly.

Inversion of the linear system (\ref{inv_decomp_0}) finally leads to
\begin{equation}
 D(j,k) = (-1)^{j+k} {j+k \choose 2k} \,,
\end{equation}
which can also be written
\begin{equation}
 D_L = U_L(\ell/2) \,,
 \label{DL0}
\end{equation}
where $U_k(x)$ is the $k$th order Chebyshev polynomial of the second
kind, $U_L(\cos\theta)={\sin(L+1)\theta\over\sin\theta}$.

The total number of states is
\begin{equation}
 \sum_{j=0}^{N_2} d_{2j} = {N \choose N/2} \,.
\end{equation}
One should also note the sum rule
\begin{equation}
 \sum_{j=0}^{N_2} d_{2j} D_{2j} = \ell^N
\end{equation}
which expresses the fact that there are $\ell$ degrees of freedom living on
each site.

The representation theory of the 0BTL algebra is of course well-known, as are
the links to the XXZ spin chain \cite{N}. For generic values of $n$, the
irreducible representations are labeled by a single integer $L=0,2,\ldots,N$
which counts the number of non-contractible (or ``through'') lines, and have
dimension equal to the multiplicity of the spin ${L\over 2}$ representation in
a chain of $N$ spins $1/2$. This dimension is easily seen to be $d_L$ of
(\ref{dL0}). Meanwhile, $D_L$ is a q-dimension for the corresponding commutant (the quantum algebra \uqsl2 with $q+q^{-1}=l$.

\subsection{The odd and dilute cases}

Although we have showed how to obtain $d_L$ and $D_L$ only for a somewhat
particular case (with $N$ even, and for a fully-packed model of loops) the
results hold true more generally.

That the expressions (\ref{dL0}) for $d_L$ and (\ref{DL0}) for $D_L$ are
correct also for $N$ (and hence $L$) odd is actually obvious from the way
they are derived. As an additional check, note that the sumrule for the
odd case indeed becomes
\begin{equation}
 \sum_{j=0}^{\lfloor N/2 \rfloor} d_{2j+1} D_{2j+1} = \ell^N \,.
\end{equation}

For the dilute case, where not all of the $N$ sites need sustain a
link, the number of states $d_L$ obviously changes. To see how, let
now $\tilde{z}$ be the weight of {\em one} site (recall that $z$ was
previously defined as the weight of two sites). Let
$\tilde{f}(\tilde{z})$ be the generating function of a state
consisting only of arcs and empty sites. If such a state is non-empty,
it contains a first site which can either be empty or occupied by an
arc.  In the former case, the remainder of the state is again an arc
state, and in the latter the leftmost arc divides space into two parts which
can each sustain an independent arc state. We have therefore
$\tilde{f}(\tilde{z}) = 1 + \tilde{z} \tilde{f}(\tilde{z}) +
                            \tilde{z}^2 \tilde{f}(\tilde{z})^2$
with regular solution
\begin{equation}
 \tilde{f}(\tilde{z}) =
 \frac{1-\tilde{z}-\sqrt{1-2\tilde{z}-3\tilde{z}^2}}{2 \tilde{z}^2} =
 \sum_{j=0}^\infty \left( \sum_{k=1}^{j+2} 
 \frac{3^{j+2-k} 2^{-(j+2)} (2k-2)!}
 {(k-1)! \, (j+2-k)! \, (2k-j-2)!} \right) \tilde{z}^j
\end{equation}
The number inside the parenthesis is known in combinatorics as the
$j$th Motzkin number \cite{Motzkin}. The generating function for
dilute states with $L$ strings is then
\begin{equation}
 \tilde{z}^L \tilde{f}(\tilde{z})^{L+1} =
 \sum_{j=L}^\infty d^{\rm dil}_L(N=j) \tilde{z}^j
\end{equation}
and defines the dilute dimensions $d^{\rm dil}_L(N)$.

We now claim that the dimensions $D_L$ of (\ref{DL0}) are unchanged in
the dilute case. While this can be seen be repeating the construction
of invariant reduced states, we shall opt instead for a simple check.
Indeed, the sumrule with $d^{\rm dil}_L$ as above and $D_L$ taken from
(\ref{DL0}) becomes
\begin{equation}
 \sum_{L=0}^N d^{\rm dil}_L D_L = (\ell + 1)^N
\end{equation}
where we note that the sum is now over both parities of $L$. This result
is exactly as expected, since the number of degrees of freedom living at
each site must indeed be $\ell+1$ (with $\ell$ coming from the loops, and
$1$ coming from the possibility of the site being empty).

In the 1BTL and 2BTL cases treated below, similar extensions to the odd and 
dilute cases may be worked out. However, since we have just seen on the simpler
0BTL example that the dimensions $D_L$---our main concern---are unchanged, we
shall not go through this exercise in what follows.

\section{One-boundary case}

\subsection{Algebraic structure}
\label{algstruc1BTL}

In the one-boundary case, contractible loops touching at least once the left
boundary receive a weight $n_{\rm l}$ which is different from the weight $n$
of a bulk loop. Coding this algebraically requires the introduction of an
additional generator $b_{\rm l}$ acting on the (left) boundary, such that
\begin{eqnarray}
 b_{\rm l}^2 &=& b_{\rm l} \nonumber \\
 e_1 b_{\rm l} e_1 &=& n_{\rm l} e_1 \nonumber \\
 \left[b_{\rm l},e_i \right] &=& 0 \mbox{ for $i=2,3,\ldots,N-1$}
 \label{TL1}
\end{eqnarray}
These relations, with (\ref{TL0}), define the one-boundary Temperley-Lieb
(1BTL) algebra \cite{MS}.

\begin{figure}
  \begin{center}
    \includegraphics[scale=0.4]{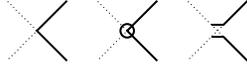}
  \end{center}
  \caption{Operators acting on the left boundary. From left to right: identity,
  blob, fork.}
  \label{fig:vertices1}
\end{figure}

Graphically, the action of $b_{\rm l}$ can be depicted by adding a
{\em blob} (shown in the following figures as a circle) to the link
that touches the boundary. This is illustrated in Fig.~\ref{fig:vertices1}.
The first relation in (\ref{TL1}) means
that all the anchoring points of a boundary touching loop, except the
last one, can be taken away. The third relation and (\ref{TL0}) allow
to deform and diminish the size of a boundary loop (while keeping it
glued to one of its anchoring points on the boundary), and when it has
reached its minimal possible size it can be taken away and replaced by
the weight $n_{\rm l}$ due to the second relation of (\ref{TL1}).

In part of the literature (see \cite{Gier02} for an example) the
algebra (\ref{TL1}) is normalized differently, and the boundary
generator is depicted as a {\em fork} rather than a blob (see
Fig.~\ref{fig:vertices1}). This means that where a loop touches the
boundary it is cut, and the two ends are attached to the boundary.
This convention allows to interpret a loop touching the boundary $k$
times as a collection of $k$ half loops with end points on the
boundary. These half loops can subsequently be pulled apart, something
which is not possible in the blob picture.  In this paper we shall
work exclusively in the blob picture, but we believe that all our
results can equivalently be derived and interpreted in the fork
picture.

The transfer matrix can be taken as
\begin{equation}
 T = \left( \prod_{j=1}^{N/2-1} (I + e_{2j}) \right)
     \left( \prod_{j=1}^{N/2} (I + e_{2j-1}) \right) (\lambda_{\rm l} I + b_{\rm l})
 \label{TM1}
\end{equation}
where a non-zero value of $\lambda_{\rm l}$ would mean that with some
probability a loop may come close to the boundary without actually
touching it. We will mostly set $\lambda_{\rm l}=0$ in what follows. This has the
advantage of reducing the dimension of the space on which $T$ acts,
since then the leftmost link in any (reduced) state may be taken to be
blobbed. The algebraic results for the case $\lambda_{\rm l}\neq 0$ are simply related to those for the 
case $\lambda_{\rm l}=0$, and we shall discuss them in due course.

\subsection{States and transfer matrix decomposition}
\label{sec:states1}

The states of the transfer matrix are as in the 0BTL case, except that links
which are exposed to the boundary (i.e., which are not to the right of the
leftmost string) may be blobbed. Also, any link touching the leftmost site
($i=1$) is necessarily blobbed, since we have taken $\lambda_{\rm l}=0$ in
(\ref{TM1}).

\begin{figure}
  \begin{center}
    \includegraphics[scale=0.4]{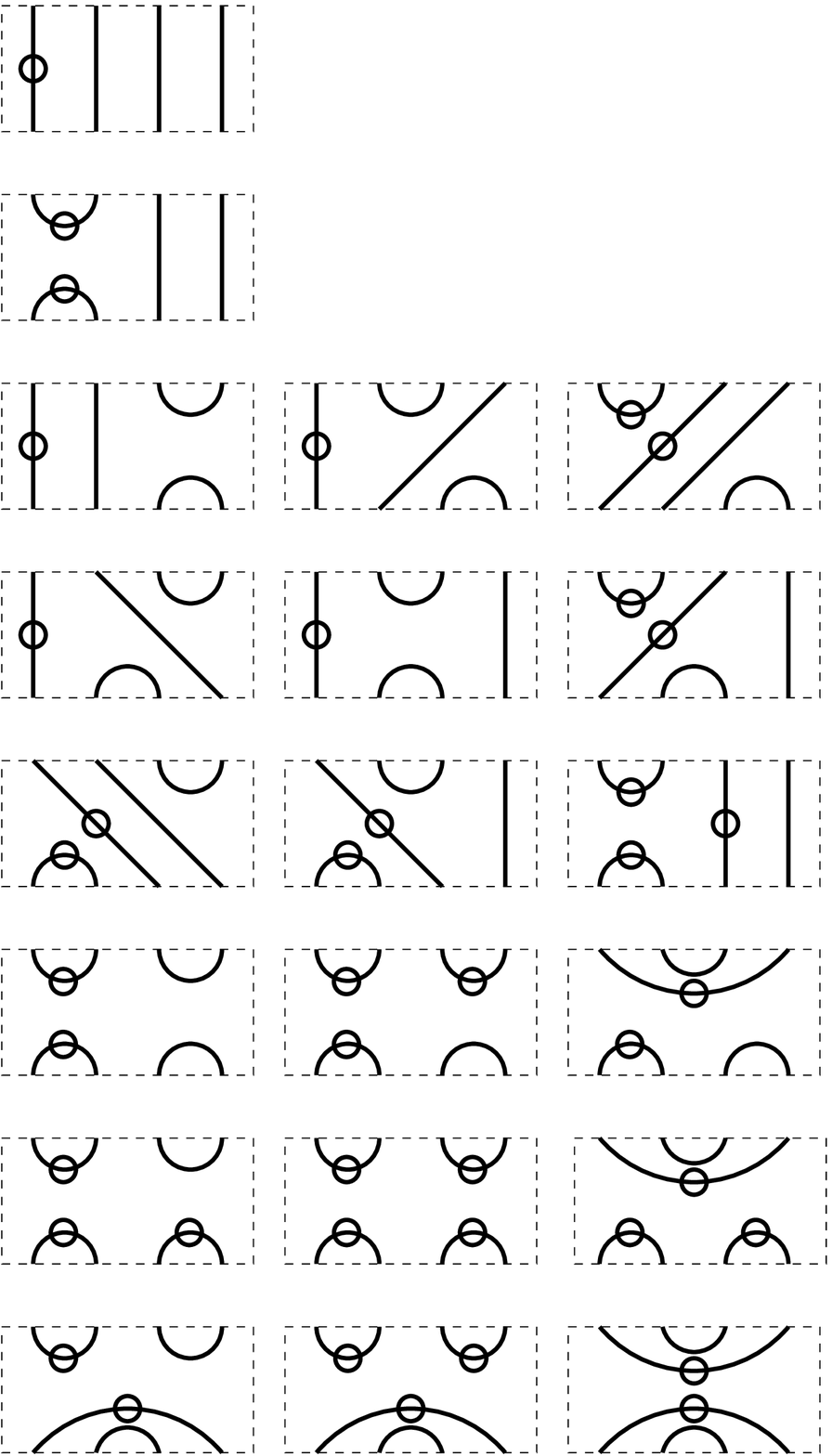}
  \end{center}
  \caption{List of all 1BTL states on $N=4$ strands (with $\lambda_{\rm l}=0$). Each
    row corresponds to a definite sector of the transfer matrix.}
  \label{fig:states1}
\end{figure}

\begin{figure}
  \begin{center}
    \includegraphics[scale=0.4]{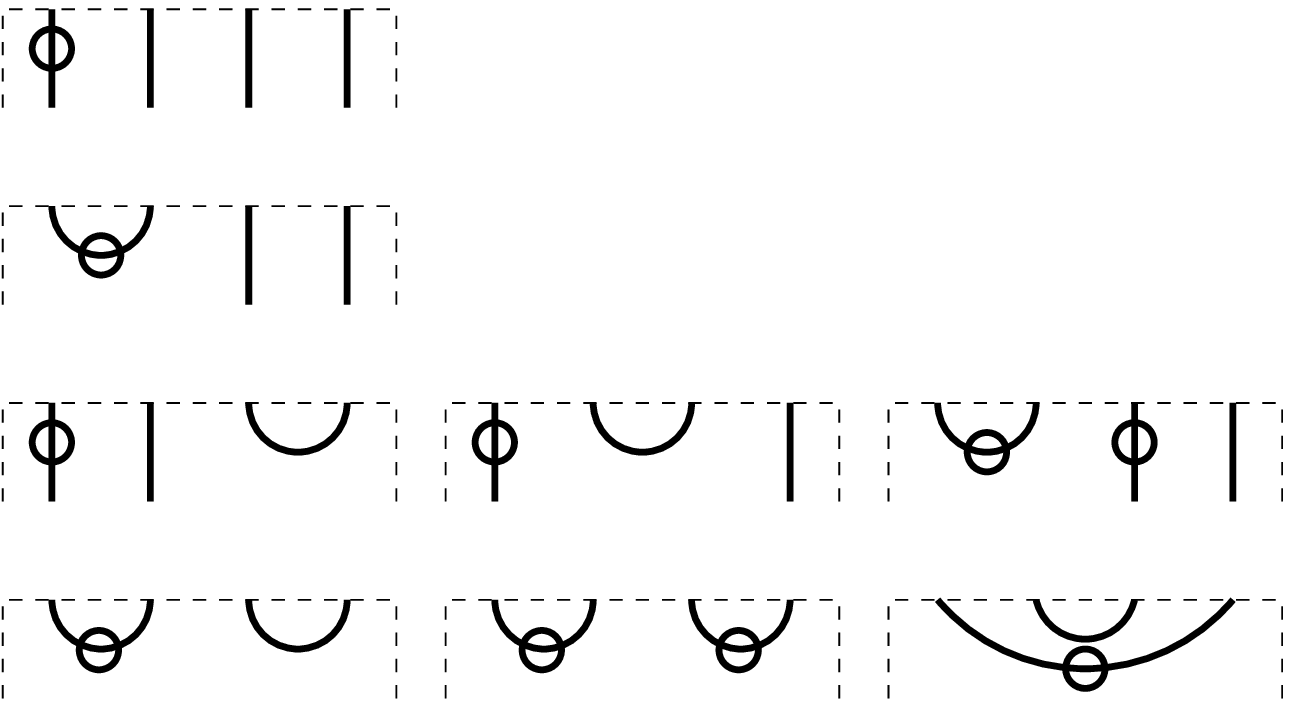}
  \end{center}
  \caption{List of all 1BTL reduced states on $N=4$ strands (with
    $\lambda_{\rm l}=0$). Each row corresponds to a definite sector of the transfer
    matrix.}
  \label{fig:red_states1}
\end{figure}

The states for $N=4$ are shown in Fig.~\ref{fig:states1}, and the
corresponding reduced states are given in Fig.~\ref{fig:red_states1}. 
When a blobbed and an unblobbed link are adjoined (e.g., when
transfering, or when forming an inner product) the result is a blobbed
(restricted) link.

It should be noted that there is a slightly different way \cite{MS} of
defining the 1BTL states
by projecting the unblobbed links onto the orthogonal complement of
the blobbed ones, i.e., associating them with $1-b_{\rm l}$. These
links can be decorated with another symbol (taken to be a small square
in Ref.~\cite{MS}).
A squared loop now comes with a weight $n-n_{\rm l}$, and when a
blobbed and a squared link are adjoined the result is zero by
orthogonality. Clearly, the two constructions are completely
equivalent. We shall not use the alternate definition in this paper.

The decomposition of the transfer matrix into sectors (blocks) takes
place exactly as in the 0BTL case, with one important addition.
Namely, once the number of strings $s=2j$ has been fixed, the blocks
$T_j$ are blockwise $2 \times 2$ lower triangular with respect to the
blobbing status of the leftmost string. Indeed, acting by $b_{\rm l}$
can blob the leftmost string, but a string---{\em qua} a conserved
object with respect to $T_j$---cannot subsequently be unblobbed. The
elementary blocks are therefore $T_j^{\rm b}$ and $T_j^{\rm u}$, where
the superscript indicates the blobbing status (b for blobbed, and u
for unblobbed) of the leftmost string. With $\lambda_{\rm l}=0$, there
is no unblobbed sector with $j=N/2$, and by convention the sector with
$j=0$ is unblobbed (for any $\lambda_{\rm l}$). In
Figs.~\ref{fig:states1}--\ref{fig:red_states1}, the second rows give
the unique unblobbed state with $j=1$.

\subsection{The dimensions $d_L^\alpha$}

Let us now determine the number of reduced states in the various sectors of the
transfer matrix. Recall that (\ref{gen_f}) gives the generating function
$f(z)$ of a collection of unblobbed arcs. We shall also need the generating
function $e(z)$ of arc states, where exterior arcs are allowed (but not
required) to be blobbed, and all interior arcs are unblobbed. Such a state is
either empty, or it has a leftmost arc (which may or may not be blobbed) which
divides the space into two parts (inside the arc and to its right). On the
inside is a collection of unblobbed arcs (i.e., a factor of $f(z)$), and on
the right is another factor of $e(z)$. Thus, $e(z) = 1 + 2 z f(z) e(z)$, or
\begin{equation}
 e(z) = \frac{1}{\sqrt{1-4z}} = \sum_{j=0}^\infty {2j \choose j} \, z^j \,.
\label{gen_e}
\end{equation}

Consider now first the case of $\lambda_{\rm l} \neq 0$ in (\ref{TM1}), where the
leftmost object may have any blobbing status. If the left string is
required to be unblobbed, the generating function in the sector with $2k$
strings is
\begin{equation}
 e(z) z^k f(z)^{2k} = \sum_{j=k}^\infty {2j \choose j-k} z^j \,,
\end{equation}
and the result is the same if the left string is required to be blobbed.
Thus,
\begin{equation}
 d^{\rm u}_L = d^{\rm b}_L = {N \choose (N-L)/2}
 \qquad (\mbox{For }\lambda_{\rm l} \neq 0)
 \label{d_lambda_not_zero}
\end{equation}
This is a well-known result for the 1BTL algebra \cite{MS}. Note in particular
that the total number of states is
\begin{equation}
 \sum_{j=0}^{N_2} d^{\rm u}_{2j} + \sum_{j=1}^{N_2} d^{\rm b}_{2j} = 2^N
 \,.
\end{equation}

Consider next the case of $\lambda_{\rm l}=0$, the convention that we shall use below,
and for which Figs.~\ref{fig:states1}--\ref{fig:red_states1} apply. In the
blobbed sector we must distinguish the case where the leftmost object is a
string or an arc. This gives
\begin{equation}
 z^k f(z)^{2k} + z^{k+1} e(z) f(z)^{2k+1} =
 \sum_{j=k}^\infty {2j-1 \choose j-k} z^j \,.
\end{equation}
Meanwhile, for the unblobbed sector, the leftmost object must be a blobbed
arc. The generating function then reads
\begin{equation}
 z^{k+1} e(z) f(z)^{2k+1} =
 \sum_{j=k}^\infty {2j-1 \choose j-k-1} z^j \,.
\end{equation}
To summarize, we have shown that there are two different types of irreducible
representations (unblobbed and blobbed), obtained by acting with the algebra
on reduced states, of dimensions
\begin{equation}
 d^{\rm u}_L = {N-1 \choose (N-L)/2-1} \,, \qquad
 d^{\rm b}_L = {N-1 \choose (N-L)/2}
 \qquad (\mbox{For }\lambda_{\rm l} = 0)
 \label{d_lambda_zero}
\end{equation}
The total number of states is now
\begin{equation}
 \sum_{j=0}^{N_2-1} d^{\rm u}_{2j} + \sum_{j=1}^{N_2} d^{\rm b}_{2j} = 2^{N-1}
 \,.
\end{equation}

Note that in the papers \cite{NRG,N}, the whole algebra is considered,
corresponding to our case $\lambda_{\rm l} \neq 0$. We shall however continue to study
the simpler quotient ($\lambda_{\rm l}=0$) below.

\subsection{The dimensions $D_L^\alpha$}

In what follows, Greek letters $\alpha$ and $\gamma$ are sector labels
which can designate the blobbed ($\alpha = {\rm b}$) or the unblobbed
($\alpha = {\rm u}$) sector. The character $K_j^\alpha$ corresponding
to a transfer matrix block $T_j^\alpha$ is defined as
\begin{equation}
 K_j^\alpha = {\rm Tr} \, \left( T_j^\alpha \right)^M \,,
 \label{Kj1}
\end{equation}
the trace being over reduced states. The constrained annulus partition
function $Z_j^\alpha$ is defined to have exactly $2j$ non-contractible
loops, of which the leftmost is blobbed or unblobbed according to the
value of $\alpha$. Recall that the number of strands $N=2N_2$ is assumed
to be even. The allowed values of $j$ in (\ref{Kj1}) are then
\begin{equation}
\begin{tabular}{lll}
 For $K_j^{\rm u}$ &:& $j=0,1,\ldots,N_2-1$ \\
 For $K_j^{\rm b}$ &:& $j=1,2,\ldots,N_2$ \\
\end{tabular}
\end{equation}
and similarly for the $Z_j^\alpha$.

The decomposition of the constrained partition functions reads
\begin{equation}
 Z_j^\alpha = \sum_{k=j}^{N_2} \left[ D^\alpha_{\rm u}(k,j) \ell^{2k} K^{\rm u}_k
            + D^\alpha_{\rm b}(k,j) \ell_{\rm l} \ell^{2k-1} K^{\rm b}_k \right] \,,
\end{equation}
where $D^\alpha_\gamma(k,j)$ are coefficients to be determined. They
define the amplitudes of the transfer matrix eigenvalues
\begin{equation}
 D^\alpha_{2j} = \sum_{i=0}^j \left[ D^\alpha_{\rm u}(j,i) \ell^{2i} +
                 D^\alpha_{\rm b}(j,i) \ell_{\rm l} \ell^{2i-1} \right] \,.
\end{equation}

As in the 0BTL case, we turn the problem upside down and consider the
inverse decomposition
\begin{equation}
 K_k^\alpha = \sum_{j=k}^{N_2} \left[ E^\alpha_{\rm u}(j,k) \frac{Z^{\rm u}_j}{\ell^{2j}} +
              E^\alpha_{\rm b}(j,k) \frac{Z^{\rm b}_j}{\ell_{\rm l} \ell^{2j-1}} \right] \,.
 \label{invdec1BTL}
\end{equation}
The coefficient $E^\alpha_\gamma(j,k)$ counts how many times each
$Z^\gamma_j$ occurs in a given trace $K_k^\alpha$. Just as in the 0BTL
case this means that we must count the number of invariant reduced
states on $2j$ strands, using $2k$ strings and $j-k$ arcs.  The
construction is the same as shown in Fig.~\ref{compstate}, with an
important modification: $\gamma$ now gives the blobbing status of the
leftmost string in Fig.~\ref{compstate}c, and $\alpha$ gives the blobbing
status of the leftmost string in the sought-for invariant reduced state. 

The number of different families of coefficients $E^\alpha_\gamma(j,k)$
to be determined is actually three rather than four. More precisely we have
\begin{eqnarray}
 E_1 &\equiv& E^{\rm u}_{\rm u} = E^{\rm b}_{\rm u} \nonumber \\
 E_2 &\equiv& E^{\rm u}_{\rm b} \\
 E_3 &\equiv& E^{\rm b}_{\rm b} \nonumber
\end{eqnarray}
defining $E_\sigma(j,k)$ for $\sigma=1,2,3$. 
The corresponding generating functions 
\begin{equation}
 E^{(k)}_\sigma(z) = \sum_{j=0}^\infty E_\sigma(j,k) z^j
\end{equation}
can then be written down in terms of those of states of unblobbed arcs $f(z)$
[see (\ref{gen_f})], and of arc states where any exterior arc may be blobbed
$e(z)$ [see (\ref{gen_e})]. This gives:
\begin{eqnarray}
 E_1^{(k)}(z) &=& z^k e(z) f(z)^{2k} 
               = \sum_{j=k}^\infty {2j \choose j-k} z^j \nonumber \\
 E_2^{(k)}(z) &=& z^{k+1} e(z) f(z)^{2k+1} 
               = \sum_{j=k+1}^\infty {2j-1 \choose j-k-1} z^j
 \label{E1E3} \\
 E_3^{(k)}(z) &=& z^k f(z)^{2k} (1 + z e(z) f(z)) 
               = \sum_{j=k}^\infty {2j-1 \choose j-k} z^j \nonumber
\end{eqnarray}

The linear system (\ref{invdec1BTL}) with (\ref{E1E3}) can now be
inverted, leading to the result
\begin{eqnarray}
 D^{\rm u}_L &=& U_{L}(\ell/2) - \ell_{\rm l} U_{L-1}(\ell/2) \nonumber \\
 D^{\rm b}_L &=& \ell_{\rm l} U_{L-1}(\ell/2) - U_{L-2}(\ell/2)\label{cheby}
\end{eqnarray}
when expressed in terms of $U_n(x)$, the $n$th order Chebyshev
polynomial of the second kind. [Note carefully that we have here
defined $U_{n}(x) = 0$ on the right-hand side when $n < 0$, which is a
non-standard choice.]

Using the identity
\begin{equation}
 U_n(x) = 2 x U_{n-1}(x) - U_{n-2}(x)
\label{Uidentity}
\end{equation}
we get the following reductions from the 1BTL case to the 0BTL case:
\begin{eqnarray}
 \left. D^{\rm u}_L \right|_{\ell_{\rm l} = \ell} &=& -U_{L-2}(\ell/2)
 \equiv -D_{L-2} \nonumber \\
 \left. D^{\rm b}_L \right|_{\ell_{\rm l} = \ell} &=& U_{L}(\ell/2)
 \equiv D_L
\end{eqnarray}

For the sumrule giving the size of the total Hilbert space there are
two choices for the $d^{\alpha}_j$. When $\lambda_{\rm l} \neq 0$ in
(\ref{TM1}) one has (\ref{d_lambda_not_zero}), giving
\begin{equation}
 \sum_{j=0}^{N_2} d^{\rm u}_{2j} D^{\rm u}_{2j} +
 \sum_{j=1}^{N_2} d^{\rm b}_{2j} D^{\rm b}_{2j} = \ell^N \,.
\end{equation}
When $\lambda_{\rm l} = 0$ one has (\ref{d_lambda_zero}), giving instead
\begin{equation}
 \sum_{j=0}^{N_2-1} d^{\rm u}_{2j} D^{\rm u}_{2j} +
 \sum_{j=1}^{N_2} d^{\rm b}_{2j} D^{\rm b}_{2j} = \ell_{\rm l} \ell^{N-1} \,.
\end{equation}
This latter result expresses clearly that the number of degrees of freedom
on the first strand has been reduced from $\ell$ to $\ell_{\rm l}$. These
two sumrules are also an important check of the fact that the dimensions
$D^\alpha_L$ depend only on the number of non-contractible lines, and not
on how one restricts the states of the transfer matrix.

\section{Two-boundary case}

\subsection{Algebraic structure}
\label{algstruc2}

In the two-boundary case, a contractible loop has four different
possibilities: it can touch none of the boundaries, only the left one,
only the right one, or touch both boundaries. The corresponding weights
have already been defined in (\ref{loopweights}).

Algebraically we still have the relations (\ref{TL0}) and (\ref{TL1}).
We shall also need the analogue of (\ref{TL1}) on the right boundary:
\begin{eqnarray}
 b_{\rm r}^2 &=& b_{\rm r} \nonumber \\
 e_{N-1} b_{\rm r} e_{N-1} &=& n_{\rm r} e_{N-1} \nonumber \\
 \left[b_{\rm r},e_i \right] &=& 0 \mbox{ for $i=1,2,\ldots,N-2$}
 \label{TL1r}
\end{eqnarray}
Obviously, the time order in which a loop touches the two boundaries does
not affect its weight, so we should further set
\begin{equation}
 \left[ b_{\rm l},b_{\rm r} \right] = 0 \,.
\end{equation}
For fixed $N$, the above relations now give an infinite number of
words, since we have not yet given a prescription for how to ``dispose
of'' loops touching both boundaries.  We shall therefore need to take
the following quotient
\begin{equation}
 \left( \prod_{j=1}^{N/2} e_{2j-1} \right) b_{\rm r}
 \left( \prod_{j=1}^{N/2-1} e_{2j} \right) b_{\rm l}
 \left( \prod_{j=1}^{N/2} e_{2j-1} \right) = n_{\rm b}
 \left( \prod_{j=1}^{N/2} e_{2j-1} \right) \,.
 \label{singlequotient}
\end{equation}
Graphically, this expresses that the smallest possible loop extending
from left to right, and blobbed on both sides, can be taken out and
replaced by a factor of $n_{\rm b}$. The remaining relations then
suffice to ensure this weighting for a loop of any size and shape
touching the two boundaries.

We emphasize again that our diagrammatic conventions are in terms of
the blob picture, rather than the fork picture \cite{Gier02} which has
been discussed briefly in section \ref{algstruc1BTL}. In the fork
picture it is natural to exploit the fact that half loops can be
pulled apart to replace the single quotient (\ref{singlequotient}) by
a double quotient \cite{GN}. This is however meaningless in the blob
picture.  Note also that the sector labels (``blobbed'' and
``unblobbed'') in the blob picture (to be discussed below) are analogous
to the parity of connections to the left and right boundaries in the
fork picture (see Definition 4.2 of Ref.~\cite{GN}).
Needless to say, the two approaches had better agree on the dimensions
of the irreducible representations of 2BTL (and we shall see below that
indeed they do).

The transfer matrix can be taken as
\begin{equation}
 T = \left( \prod_{j=1}^{N/2-1} (I + e_{2j}) \right)
     (\lambda_{\rm r} I + b_{\rm r})
     \left( \prod_{j=1}^{N/2} (I + e_{2j-1}) \right)
     (\lambda_{\rm l} I + b_{\rm l})
 \label{TM2}
\end{equation}
where non-zero values of $\lambda_{\rm l}$ and $\lambda_{\rm r}$ mean
(as in the 1BTL case) that with some probability a loop may come close
to a boundary without actually touching it. We have already seen in
the 1BTL case how the choice of these parameters changes slightly the
counting of states, and the sumrules linking $d_L$ and $D_L$.

We now wish to determine with which amplitudes the eigenvalues of the
transfer matrix $T$ enter into the annulus partition function. These
amplitudes, which can also be interpreted as operator multiplicities,
depend only on the weights of non contractible loops ($\ell$,
$\ell_{\rm l}$, $\ell_{\rm r}$, and $\ell_{\rm b}$). In particular,
they are independent of the lattice structure, local occupancy
constraints (fully-packed or dilute case), and of local vertex
weights. The amplitudes are also independent of the weights of
contractible loops, as long as these are non-zero (but they do depend
on whether $n_{\rm b}=0$ or $n_{\rm b} \neq 0$, as discussed in
details below). To simplify the discussion we shall henceforth deal
with the fully-packed model only.

\subsection{States and transfer matrix decomposition}

\begin{figure}
  \begin{center}
    \includegraphics[scale=0.45]{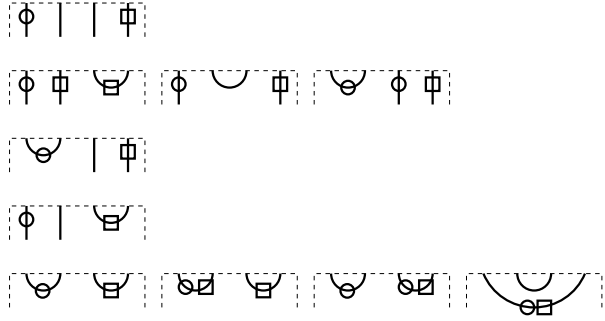}
  \end{center}
  \caption{List of all 2BTL reduced states on $N=4$ strands (with
    $\lambda_{\rm l}=\lambda_{\rm r}=0$). Each row corresponds to a
    definite sector of the transfer matrix.}
  \label{fig:red_states2}
\end{figure}

Let $T$ be the 2BTL transfer matrix (\ref{TM2}). The concept of
states, made up of links which are arcs or strings, has already been
defined in the 0BTL case.  The states on which $T$ acts are as in the
1BTL case, except that links which are exposed to the right boundary
(i.e., which are not to the left of the rightmost string) may be
blobbed by the second boundary generator $b_{\rm r}$. It is necessary
to introduce distinct symbols for the left blob (a circle) and the
right blob (a square). The reduced states for $N=4$ are shown in
Fig.~\ref{fig:red_states2}.

Let $s$ be the number of strings in a given state.  Note that $T$
cannot change the parity of $s$. We have therefore $T = T_{\rm even}
\oplus T_{\rm odd}$. As usual we assume $N=2N_2$ even, and so the
partition function
\begin{equation}
 Z^{\rm even}_{N,M} = \left \langle u | T_{\rm even}^M | v \right \rangle \,.
\label{partfunc}
\end{equation}
is constrained to an even number of non contractible lines.  Note that
the loop weight $\ell_{\rm b}$ cannot appear in $Z^{\rm even}$ (and
similarly $n_{\rm b}$ cannot appear in $Z^{\rm odd}$). Henceforth we
shall drop the epithet ``even''.

Only the leftmost string, and the arcs to its left, can be blobbed by
$b_{\rm l}$. Similarly, only the rightmost string, and the arcs to its
right, can be blobbed by $b_{\rm r}$. When $s=0$, at most one arc can
be blobbed on both sides (i.e., by $b_{\rm l} b_{\rm r}$). If there is
such a doubly blobbed arc, any arc to its left (resp.\ right) cannot
be blobbed by $b_{\rm r}$ (resp.\ $b_{\rm l}$). When $n_{\rm b} = 0$
doubly blobbed arcs are forbidden, so in the sequel we shall have to
distinguish between the cases $n_{\rm b} = 0$ and $n_{\rm b} \neq 0$.
(To compute $Z_{\rm odd}$ we would similarly have to distinguish
between $\ell_{\rm b} = 0$ and $\ell_{\rm b} \neq 0$, when $s=1$).

The states can be ordered as follows: First we sort the states
according to a decreasing number of strings $s$. For fixed $s>0$, we
place first the states in which the leftmost and rightmost strings are
both unblobbed (henceforth called uu states), then states in which
only the rightmost string is blobbed (ub states), then states in which
only the leftmost string is blobbed (bu states), and finally states in
which both the leftmost and rightmost strings are blobbed (bb states).
Having done this, we finally group together states (with fixed $s$ and
fixed blobbing (uu, ub, bu, or bb) of the outermost strings) that possess
an equal lower-half reduced state.  With this ordering of the states,
$T$ has a lower block-triagonal structure, with each block
corresponding to a group of states as defined above.

The blocks on the diagonal of $T$ are denoted $T_j^{\alpha \beta}$,
where $j=s/2$ is the number of {\em pairs} of strings, and the indices
$\alpha,\beta={\rm u}$ or ${\rm b}$. Note that the blocks $T_j^{\alpha
  \beta}$ can be constructed in terms of the reduced states.

\subsection{The dimensions $D_L^{\alpha \beta}$}

In what follows $\alpha,\beta,\gamma,\delta = {\rm u}$ or ${\rm b}$
are sector labels.  For each transfer matrix block $T_j^{\alpha
  \beta}$ we define the corresponding character as
\begin{equation}
 K_j^{\alpha \beta} = {\rm Tr}\, \left( T_j^{\alpha \beta} \right)^M,
 \label{charK}
\end{equation}
where as usual the trace is over reduced states.  Also, let
$Z_j^{\alpha \beta}$ be the annulus partition function constrained to
have exactly $j$ non contractible loops, of which the leftmost (resp.\
rightmost) has blobbing status $\alpha$ (resp.\ $\beta$).  For
example, $Z_j^{\rm ub}$ consists of the terms in the full partition
function $Z_{N,M}$ whose dependence on $\ell$, $\ell_{\rm l}$,
$\ell_{\rm r}$ is precisely $\ell^{2j-1} \ell_{\rm r}$.

The allowed values of $j$ in (\ref{charK}) are then
\begin{equation}
\begin{tabular}{lll}
 For $K_j^{\rm uu}$ &:& $j=0,1,\ldots,N_2-2$ \\ 
 For $K_j^{\rm ub}$ and $K_j^{\rm bu}$ &:& $j=1,2,\ldots,N_2-1$ \\
 For $K_j^{\rm bb}$ &:& $j=1,2,\ldots,N_2$ \\
\end{tabular}
\end{equation}
and similarly for the $Z_j^{\rm \alpha \beta}$. All other characters
and constrained partition functions are defined to be zero in order
to lighten the notation in the following formulae.

The goal is now to search for a decomposition of the form
\begin{eqnarray}
 Z_j^{\alpha \beta} &=& \sum_{k=j}^{N_2} \left[
 D^{\alpha \beta}_{\rm uu}(k,j) \ell^{2k} K^{\rm uu}_k +
 D^{\alpha \beta}_{\rm ub}(k,j) \ell^{2k-1} \ell_{\rm r} K^{\rm ub}_k \right.
 \nonumber \\
 &+& \left.
 D^{\alpha \beta}_{\rm bu}(k,j) \ell_{\rm l} \ell^{2k-1} K^{\rm bu}_k +
 D^{\alpha \beta}_{\rm bb}(k,j) \ell_{\rm l} \ell^{2k-2} \ell_{\rm r}
 K^{\rm bb}_k \right]
\label{decomp1}
\end{eqnarray}
where $D^{\alpha \beta}_{\gamma \delta}$ are coefficients to be
determined. The complete amplitude---here constructed combinatorially
monomial by monomial---then reads
\begin{equation}
 D^{\alpha \beta}_{2j} = \sum_{i=0}^j \left[
 D^{\alpha \beta}_{\rm uu}(j,i) \ell^{2i} +
 D^{\alpha \beta}_{\rm ub}(j,i) \ell^{2i-1} \ell_{\rm r} +
 D^{\alpha \beta}_{\rm bu}(j,i) \ell_{\rm l} \ell^{2i-1} +
 D^{\alpha \beta}_{\rm bb}(j,i) \ell_{\rm l} \ell^{2i-2} \ell_{\rm r} \right].
\label{decomp2}
\end{equation}

The inverse decomposition of $K_k$ in terms of $Z_j$ reads
\begin{equation}
 K_k^{\alpha \beta} = \sum_{j=k}^{N_2} \left[
 E^{\alpha \beta}_{\rm uu}(j,k) \frac{Z^{\rm uu}_j}{\ell^{2j}} +
 E^{\alpha \beta}_{\rm ub}(j,k) \frac{Z^{\rm ub}_j}{\ell^{2j-1} \ell_{\rm r}} +
 E^{\alpha \beta}_{\rm bu}(j,k) \frac{Z^{\rm bu}_j}{\ell_{\rm l} \ell^{2j-1}} +
 E^{\alpha \beta}_{\rm bb}(j,k) \frac{Z^{\rm bb}_j}
 {\ell_{\rm l} \ell^{2j-2} \ell_{\rm r}} \right]
\label{decomp3}
\end{equation}
The coefficients $E^{\alpha \beta}_{\gamma \delta}(j,k)$ counts how
many times each $Z_j^{\gamma \delta}$ occurs in a given trace
$K_k^{\alpha \beta}$, expressed in terms of invariant reduced states
on $2j$ strands, using $2k$ strings.

\begin{figure}
  \begin{center}
    \leavevmode
    \epsfysize=25mm{\epsffile{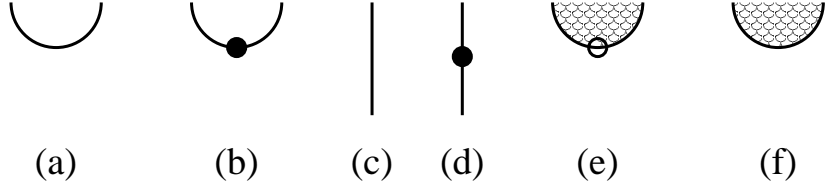}}
  \end{center}
  \protect\caption{Diagrams representing (a) an unblobbed arc, (b) a
    blobbed arc, (c) an unblobbed string, (d) a blobbed string, (e) a
    collection of arcs of which exterior arcs may (but need not) be
    blobbed, and (f) a collection of arcs of which all exterior arcs
    are unblobbed.}
  \label{diagrams}
\end{figure}

The various symmetries in the problem reduce the number of families of
coefficients $E^{\alpha \beta}_{\gamma \delta}(j,k)$ to be determined
from sixteen to six. More precisely we have:
\begin{eqnarray}
 E_1 &\equiv& E^{\rm uu}_{\rm uu} = E^{\rm ub}_{\rm uu} =
              E^{\rm bu}_{\rm uu} = E^{\rm bb}_{\rm uu} \nonumber \\
 E_2 &\equiv& E^{\rm uu}_{\rm ub} = E^{\rm uu}_{\rm bu} =
              E^{\rm bu}_{\rm ub} = E^{\rm ub}_{\rm bu} \nonumber \\
 E_3 &\equiv& E^{\rm uu}_{\rm bb} \\
 E_4 &\equiv& E^{\rm ub}_{\rm ub} = E^{\rm bu}_{\rm bu} =
              E^{\rm bb}_{\rm ub} = E^{\rm bb}_{\rm bu} \nonumber \\
 E_5 &\equiv& E^{\rm ub}_{\rm bb} = E^{\rm bu}_{\rm bb} \nonumber \\
 E_6 &\equiv& E^{\rm bb}_{\rm bb} \nonumber
\label{6genfuncs}
\end{eqnarray}
defining $E_\sigma(j,k)$ for $\sigma=1,2,\ldots,6$. To count the
corresponding number of invariant states we introduce the diagrammatic
symbols shown in Fig.~\ref{diagrams}.

\begin{figure}
  \begin{center}
    \leavevmode
    \epsfysize=60mm{\epsffile{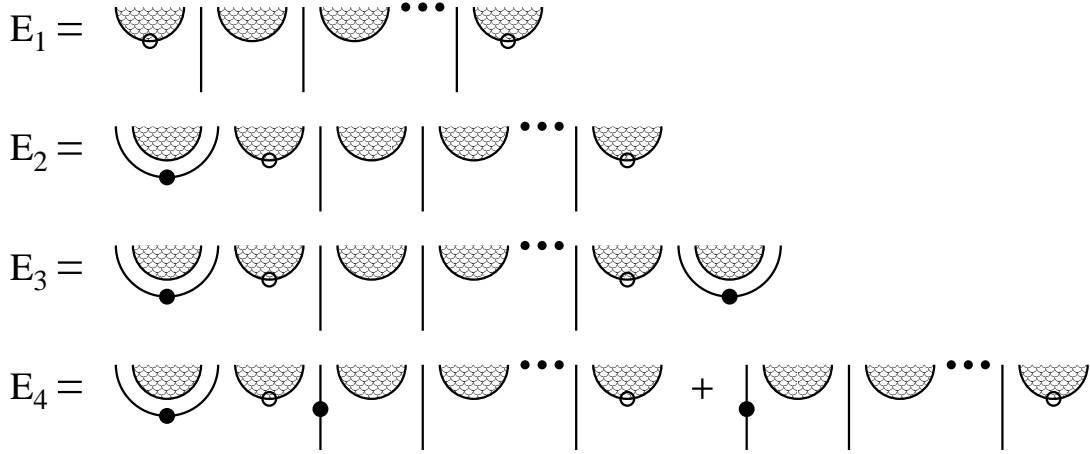}}
  \end{center}
  \protect\caption{Diagrammatics corresponding to the generating functions
    $E_\sigma(z)$ with $\sigma=1,2,3,4$.}
  \label{genfuncs}
\end{figure}

Now let
\begin{equation}
 E^{(k)}_\sigma(z) = \sum_{j=0}^\infty E_\sigma(j,k) z^j
\label{Efuncs}
\end{equation}
be the corresponding generating function, where $z$ is a formal
parameter representing the weight of an arc, or of a pair of strings.
We shall also need the generating functions $e(z)$ and $f(z)$
corresponding to the diagrams (e) and (f) in Fig.~\ref{diagrams}.  In
Eqs.~(\ref{gen_e}) and (\ref{gen_f}) these have already been found to
be
\begin{equation}
 e(z) = \frac{1}{\sqrt{1-4z}} = \sum_{j=0}^\infty {2j \choose j} \, z^j \,,
 \qquad
 f(z) = \frac{1 - \sqrt{1-4z}}{2z} = \sum_{j=0}^\infty 
 \frac{(2j)!}{j!(j+1)!} \, z^j \,.
\end{equation}
In terms of these the generating functions $E^{(k)}_\sigma(z)$ can be
written down by respecting carefully the invariance of the states.
This is shown diagramatically for for a few sample cases
($\sigma=1,2,3,4$) in Fig.~\ref{genfuncs}. In algebraic terms we have
then
\begin{eqnarray}
 E^{(k)}_1(z) &=& z^k e(z)^2 f(z)^{2k-1} \nonumber \\
 E^{(k)}_2(z) &=& z^{k+1} e(z)^2 f(z)^{2k} \nonumber \\
 E^{(k)}_3(z) &=& z^{k+2} e(z)^2 f(z)^{2k+1}
 \label{resultsE} \\
 E^{(k)}_4(z) &=& z^k e(z) f(z)^{2k-1} \big( 1 + z e(z) f(z) \big) \nonumber \\
 E^{(k)}_5(z) &=& z^{k+1} e(z) f(z)^{2k} \big( 1 + z e(z) f(z) \big) \nonumber \\
 E^{(k)}_6(z) &=& z^k f(z)^{2k-1} \big( 1 + 2 z e(z) f(z) + z^2 e(z)^2 f(z)^2 \big)
 \nonumber
\end{eqnarray}

It should be stressed that these expressions have been worked out
diagramatically by supposing the number of strings $2k>0$. But the
first three functions actually enter also into expressions with $k=0$.
It can be checked that the generating functions (\ref{resultsE}) are still
correct upon setting $k=0$, {\em provided that} we accept that no arc
is allowed to be blobbed simultaneously on the left and on the right,
i.e., provided that we set $n_{\rm b}=0$ in the notation of
(\ref{loopweights}). Let us therefore suppose that $n_{\rm b}=0$ for now,
and return to the issue of $n_{\rm b} \neq 0$ later on.

The linear system (\ref{decomp3}) with (\ref{resultsE}) can now be
inverted and brought into the form (\ref{decomp1}). The results for
the amplitudes (\ref{decomp2}) have a rather simple form
\begin{eqnarray}
 D^{\rm uu}_L &=& U_{L}(\ell/2) - (\ell_{\rm l}+\ell_{\rm r}) U_{L-1}(\ell/2)
               +  \ell_{\rm l} \ell_{\rm r} U_{L-2}(\ell/2) \nonumber \\
 D^{\rm ub}_L &=& \ell_{\rm r} U_{L-1}(\ell/2) - (1+\ell_{\rm l} \ell_{\rm r})
                  U_{L-2}(\ell/2) + \ell_{\rm l} U_{L-3}(\ell/2) \nonumber \\
 D^{\rm bu}_L &=& \ell_{\rm l} U_{L-1}(\ell/2) - (1+\ell_{\rm l} \ell_{\rm r})
                  U_{L-2}(\ell/2) + \ell_{\rm r} U_{L-3}(\ell/2)
 \label{finalD} \\
 D^{\rm bb}_L &=& \ell_{\rm l} \ell_{\rm r} U_{L-2}(\ell/2) -
                  (\ell_{\rm l}+\ell_{\rm r}) U_{L-3}(\ell/2) +
                  U_{L-4}(\ell/2) \nonumber
\end{eqnarray}
when expressed in terms of $U_n(x)$, the $n$th order Chebyshev polynomial
of the second kind. [Note again that we have defined $U_{n}(x) = 0$
on the right-hand side when $n < 0$, which is a non-standard choice.]
This is the main result of this section.

Using the identity (\ref{Uidentity}) it is easy to show that we have
reduction from the two-boundary to the one-boundary case provided that
the boundary which is non-distinguished in the latter case were
blobbed in the former. More precisely we find that
\begin{equation}
 \left. D^{\rm bb}_L \right|_{\ell_{\rm r} \to \ell} = D^{\rm b}_L \,, \qquad
 \left. D^{\rm ub}_L \right|_{\ell_{\rm r} \to \ell} = D^{\rm u}_L \,.
\end{equation}
A slightly more curious set of identities results when we remove the
distinction of an unblobbed boundary. We have then
\begin{equation}
 \left. D^{\rm bu}_L \right|_{\ell_{\rm r} \to \ell} = -D^{\rm b}_{L-2}
 \,, \qquad
 \left. D^{\rm uu}_L \right|_{\ell_{\rm r} \to \ell} = -D^{\rm u}_{L-2} \,.
\end{equation}
Finally, (\ref{Uidentity}) permits to prove the fusion identity
\begin{equation}
 \left. D^{\rm b}_2 \right|_{\ell_{\rm l} \to \ell_{\rm r}} \cdot
  D^{\rm b}_L
 = D^{\rm bb}_{L+2} + D^{\rm bb}_{L} + D^{\rm b}_{L-2}
\label{fusionidentity}
\end{equation}
where we stress that the last term on the right-hand side is a
one-boundary amplitude. More complicated fusion identities can be proved in
the same manner. The algebraic origin of these identities will be discussed in the following section.

\subsection{The dimensions $d_L^{\alpha \beta}$}

We must now compute the transfer matrix dimensions $d_L^{\alpha
  \beta}$ in the various sectors $\alpha \beta$. Recall that these are
expected to depend on the choice of $\lambda_{\rm l}$ and
$\lambda_{\rm r}$ in the transfer matrix (\ref{TM2}). Define the index
$\gamma={\rm u}$ (resp.\ $\gamma={\rm b}$) if $\lambda_{\rm l} \neq 0$
(resp.\ $\lambda_{\rm l} = 0$). Similarly define the index
$\delta={\rm u}$ or ${\rm b}$ in terms of $\lambda_{\rm r}$.
A renewed inspection of the diagrammatics of Fig.~\ref{genfuncs} then
reveals that we have simply
\begin{equation}
 d_L^{\alpha \beta}(\gamma \delta) =
 E^{\alpha \beta}_{\gamma \delta}(N/2,L/2) \,,
\end{equation}
where the coefficients $E^{\alpha \beta}_{\gamma \delta}(j,k)$ are those
of the preceding subsection; see Eq.~(\ref{6genfuncs}).

Consider in particular the case of the full 2BTL algebra (with
$\lambda_{\rm l} \neq 0$ and $\lambda_{\rm r} \neq 0$) for which
(\ref{6genfuncs}) gives $d^{\alpha \beta}_L = E^{\alpha \beta}_{\rm
  uu} = E_1$ independently of $\alpha$ and $\beta$. The coefficient of
$z^{N/2}$ in the development of $E_1^{L/2}(z)$, given explicitly in
(\ref{resultsE}), then yields exactly the dimension of an irreducible
representation of the 2BTL algebra, in complete agreement with
Proposition 4.1 of Ref.~\cite{GN}. As already announced in
section~\ref{algstruc2}, this furnishes a highly non-trivial check that
the blob and fork pictures lead to equivalent results.

\begin{figure}
  \begin{center}
    \leavevmode
    \epsfysize=80mm{\epsffile{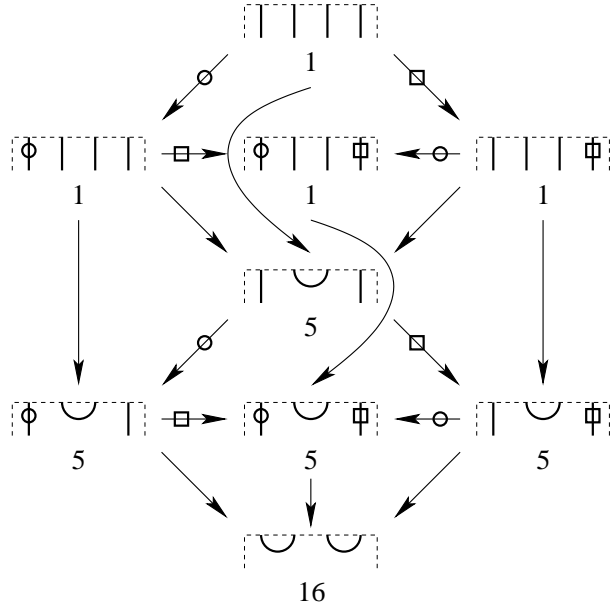}}
  \end{center}
  \protect\caption{Organization of transfer matrix blocks (irreducible
    modules) on $N=4$ strands for the full 2BTL algebra ($\lambda_{\rm
      l} \neq 0$, $\lambda_{\rm r} \neq 0$ and $n_{\rm b} \neq 0$).
    For each block we show its dimension and one representative state.
    Undecorated arrows denote transitions induced by the TL generators
    $e_i$. Arrows decorated by a circle (resp.\ a square) denote
    transitions induced by the left (resp.\ right) boundary generator
    $b_{\rm l}$ (resp.\ $b_{\rm r}$).}
  \label{organization1}
\end{figure}

\begin{figure}
  \begin{center}
    \leavevmode
    \epsfysize=45mm{\epsffile{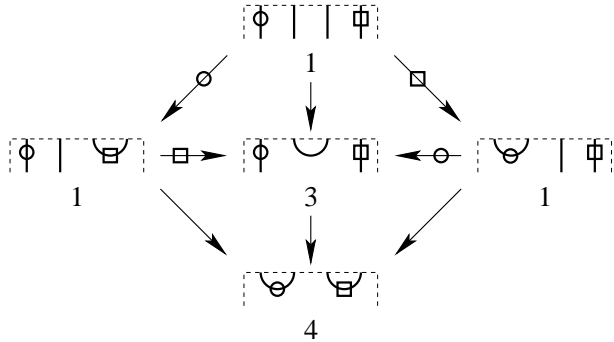}}
  \end{center}
  \protect\caption{Same as Fig.~\ref{organization1}, but for the
    restricted 2BTL algebra ($\lambda_{\rm l} = \lambda_{\rm r} = 0$
    and $n_{\rm b} \neq 0$).}
  \label{organization2}
\end{figure}

Our interpretation of the irreducible modules (transfer matrix blocks)
is however quite different from that of the fork picture \cite{GN}. To
make this point clear, we invite the reader to compare Figure 5 of
Ref.~\cite{GN} with our Fig.~\ref{organization1}. Both show the
organization of the irreducible modules and their dimensions for a
system on $N=4$ strands. In Fig.~\ref{organization2} we give the
similar picture for the restricted 2BTL algebra ($\lambda_{\rm
  l}=\lambda_{\rm r}=0$).

The following sumrules on the dimensions are readily established:
\begin{eqnarray}
 \gamma \delta = {\rm uu} &:&
 \sum_{j=0}^{N_2} d_{2j}^{\rm uu} D_{2j}^{\rm uu} +
 \sum_{j=0}^{N_2} d_{2j}^{\rm ub} D_{2j}^{\rm ub} +
 \sum_{j=0}^{N_2} d_{2j}^{\rm bu} D_{2j}^{\rm bu} +
 \sum_{j=0}^{N_2} d_{2j}^{\rm bb} D_{2j}^{\rm bb} = \ell^N
 \nonumber \\
 \gamma \delta = {\rm ub} &:&
 \sum_{j=0}^{N_2-1} d_{2j}^{\rm uu} D_{2j}^{\rm uu} +
 \sum_{j=1}^{N_2} d_{2j}^{\rm ub} D_{2j}^{\rm ub} +
 \sum_{j=1}^{N_2-1} d_{2j}^{\rm bu} D_{2j}^{\rm bu} +
 \sum_{j=1}^{N_2} d_{2j}^{\rm bb} D_{2j}^{\rm bb} = \ell^{N-1} \ell_{\rm r}
 \nonumber \\
 \gamma \delta = {\rm bu} &:&
 \sum_{j=0}^{N_2-1} d_{2j}^{\rm uu} D_{2j}^{\rm uu} +
 \sum_{j=1}^{N_2-1} d_{2j}^{\rm ub} D_{2j}^{\rm ub} +
 \sum_{j=1}^{N_2} d_{2j}^{\rm bu} D_{2j}^{\rm bu} +
 \sum_{j=1}^{N_2} d_{2j}^{\rm bb} D_{2j}^{\rm bb} = \ell_{\rm l} \ell^{N-1}
 \nonumber \\
 \gamma \delta = {\rm bb} &:&
 \sum_{j=0}^{N_2-2} d_{2j}^{\rm uu} D_{2j}^{\rm uu} +
 \sum_{j=1}^{N_2-1} d_{2j}^{\rm ub} D_{2j}^{\rm ub} +
 \sum_{j=1}^{N_2-1} d_{2j}^{\rm bu} D_{2j}^{\rm bu} +
 \sum_{j=1}^{N_2} d_{2j}^{\rm bb} D_{2j}^{\rm bb} = \ell_{\rm l} \ell^{N-2} \ell_{\rm r}
 \label{2B_sumrules}
\end{eqnarray}
where we draw the reader's attention to the summation limits.

The total number of states, for instance when $\lambda_{\rm
  l}=\lambda_{\rm r}=0$, is
\begin{equation}
 \sum_{j=0}^{N_2-2} d_{2j}^{\rm uu} +
 \sum_{j=1}^{N_2-1} d_{2j}^{\rm ub} +
 \sum_{j=1}^{N_2-1} d_{2j}^{\rm bu} +
 \sum_{j=1}^{N_2} d_{2j}^{\rm bb} =
 \sum_{k=0}^{N_2-1} \frac{2k+1}{N_2+k} {2N_2-2 \choose N_2-k-1}
 {2(k+1) \choose 2}
\end{equation}
The numerical values of this for $N=2,4,6,\ldots$ are the following:
$1,7,35,162,723,3158,\ldots$.

\subsection{Allowing doubly blobbed arcs}
\label{doubly_blobbed}

Removing the restriction $n_{\rm b} = 0$ changes our results, since now
a single arc is allowed to touch both boundaries. This quests us to
recompute $E_1(z)$, $E_2(z)$ and $E_3(z)$ for the case of $k=0$ strings.
Skipping the details, the results are simply:
\begin{equation}
 E^{(0)}_1(z) = \frac{1}{1-4z}, \quad
 E^{(0)}_2(z) = \frac{2z}{1-4z}, \quad
 E^{(0)}_3(z) = \frac{z}{1-4z} \qquad \mbox{for $k=0$ and $n_{\rm b} \neq 0$} \,.
\end{equation}
Inverting again (\ref{decomp3}) leads to exactly the same result
(\ref{finalD}) as before, {\em with one exception}: the
former result $D^{\rm bb}_2 = \ell_{\rm l} \ell_{\rm r}$ gets replaced
by
\begin{equation}
 D^{\rm bb}_2 = \ell_{\rm l} \ell_{\rm r} - 1 \qquad
 \mbox{for } n_{\rm b} \neq 0 \,.
\end{equation}
Despite of this modification of the $D_L^{\alpha \beta}$, the sumrules
(\ref{2B_sumrules}) still hold true.

\section{The algebraic significance  of  the  $D$ numbers}

The $d$ numbers are clearly dimensions of (generically) irreducible
representations of the various versions of the Temperley Lieb algebra.
As discussed briefly in our first paper \cite{JS}, the $D$ numbers
also have an interesting algebraic interpretation which we would like
to elaborate now.

The general idea is to interpret the loop configurations in the model
(either in the dense or dilute case) as the graphical expansion of an
O($n$) type model of vector spins, each spin located at a lattice
vertex. This generalizes the well-known construction of the bulk
O($n$) model to a model with boundary. Bulk spins take values in a
space of dimension $n$, hence bulk loops get a fugacity $n$. The rules
for the one-boundary loop model for instance correspond now to
constraining boundary spins to take values in a smaller space of
dimension $n_{\rm l}$. Blobbed links then indicate that the spins have
been restricted to the smaller space, while unblobbed links indicate
no restriction.  When a blobbed and an unblobbed link are adjoined
(e.g., when transfering, or when forming an inner product) the result
is a blobbed (restricted) link.

In a model with genuine O($n$) symmetry, the $D$ numbers should be the
dimensions of irreducible representations of the symmetry group, since
they would then be multiplicities of eigenvalues.  For instance, the
multiplicity $n$ should come with the eigenvalue associated with the
order parameter, {\em etc}. The slightly more formal structure behind this
is a decomposition of the Hilbert space of such an O($n$) model into a
sum of products of representations of the symmetry times
representations of the commutant. The question then is to adapt these
ideas to the case of loop models.

While loop models for arbitrary values of the parameter $n$ are
difficult to fully understand in terms of symmetry, such an
understanding was recently achieved \cite{RS07-1} for integer
(positive or negative) values of $n$. We propose here to interpret the
1BTL and 2BTL algebra and the $D$ numbers within this framework.

We start with some reminders. A cursory look at the $D$ numbers in
the ordinary TL case reveals quantities which, for $n$ integer, do not
reproduce dimensions of irreducibles of O($n$), but allow for
considerably higher degeneracies! The point is that the dense model
has a much larger symmetry than O($n$). Since loop crossings are not
allowed, one can expect at least the symmetry U($n$); a closer look
shows that this symmetry is in a certain sense realized ``locally'',
giving rise to a yet larger symmetry algebra dubbed ${\cal A}_{n}$ in
\cite{RS07-1}.
 
A convenient way to understand the enlarged symmetry is to give
explicit expressions for all its generators. This was done in
\cite{RS07-1} and we start by recalling some results from this
reference.
 
We first consider the case $n$ a positive integer, and a Hilbert space
made of $N=2N_2$ sites labelled $i=1, \ldots, N$, with an
$n$-dimensional complex vector space $V_i= {\bf C}^n$ at each site.
The states can be represented using oscillator operators $b_i^a$,
$b_{ia}^\dagger$ for $i$ even, $\overline{b}_{ia}$,
$\overline{b}_i^{a\dagger}$ for $i$ odd, with commutation relations
$[b_i^a,b_{jb}^\dagger]= \delta_{ij}\delta_b^a$ (where $a,b=1,2,\ldots,
n$), and similarly for $i$ odd. The destruction operators $b_i^a$,
$\overline{b}_{ia}$ destroy the vacuum state, the daggers indicate the
adjoint, and the spaces $V_i$ are defined by the constraints
\begin{eqnarray}
 b_{ia}^\dagger b_i^a&=& 1 \quad (i\hbox{ even}),\label{constr1}\\
 \overline{b}_i^{a\dagger} \overline{b}_{ia} &=& 1 \quad (i\hbox{
 odd})\label{constr2}
\end{eqnarray}
of one boson per site (we use the summation convention for repeated
indices of the same type as $a$). We define the generators of SU($n$)
(or in fact of gl$_n$) acting in the spaces $V_i$ by
$J_{ia}^b=b_{ia}^\dagger b_i^b$ for $i$ even,
$J_{ia}^b=-\overline{b}_i^{b\dagger} \overline{b}_{ia}$ for $i$ odd,
and the commutation relations among the $J_i$s (for each $i$) are
$i$-independent. Hence the global gl$_n$ algebra, defined by its
generators $J_a^b=\sum_i J_{ia}^b$, acts in the tensor product
$V=\otimes_{i=1}^{N} V_i$ of copies of the fundamental representation
of gl$_m$ on even sites, alternating with its dual on odd sites. Though
the U($1$) subalgebra of gl$_n$ generated by $J_a^a$ acts trivially on
the chain (and by a scalar on each site), it is often notationally
convenient not to subtract this trace from the generators $J_a^b$.

The invariant nearest-neighbor coupling in the chain is
unique, up to additive and multiplicative constants. It is the usual
``Heisenberg coupling'' of magnetism, and can be written in terms of
operators $e_i$, defined explicitly as
\be
 e_i=\left\{\begin{array} {rl}
    \overline{b}_{i+1}^{a\dagger}b_{ia}^\dagger
    b_i^b \overline{b}_{i+1,b},&\hbox{ $i$ even,}\\
    \overline{b}_i^{a\dagger}b_{i+1,a}^\dagger b_{i+1}^b
    \overline{b}_{ib},&\hbox{ $i$ odd.}
    \end{array}\right.
\ee
The $e_i$'s are Hermitian, $e_i^\dagger=e_i$. Acting in the
constrained space $V$, they satisfy the usual TL relations (\ref{TL0}).

\subsection{Zero-boundary case}

\begin{figure}
  \begin{center}
    \includegraphics[scale=0.4]{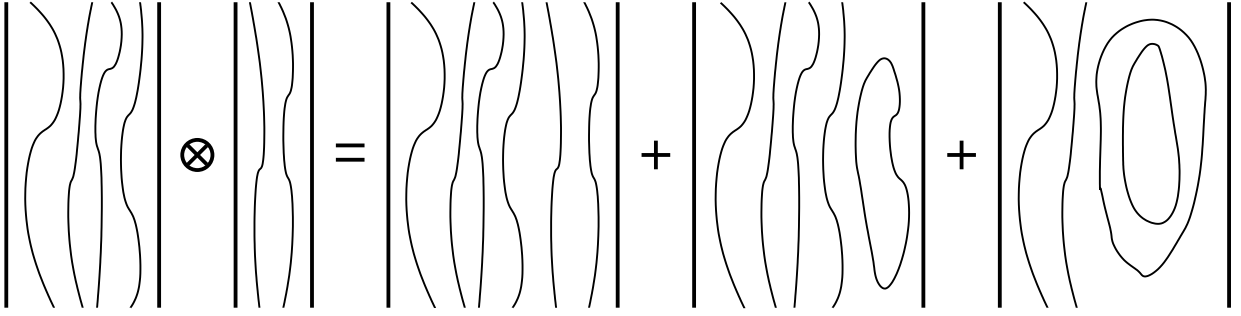}
  \end{center}
  \caption{Fusion in the ordinary TL model.}
  \label{fig:trick}
\end{figure}

In the case of ordinary (0BTL) boundary conditions, the numbers $D_{L}$ are
the dimensions $D_L$ of the $L$th representation of the commutant of
TL$_{N}(n)$ in $V$. These numbers can be found inductively, by adding
another pair of non-contractible dots to the end of a sequence. This
leads easily to the recurrence relation\footnote{In some of our other
  papers, labels for the representations are $SU(2)$ spins, and thus
  one half of those used here.}
\be
  D_2 D_L=D_{L+2}+D_L+D_{L-2}.
\ee
which is illustrated pictorially in Fig.~\ref{fig:trick}.

Also, it is clear that the initial values read $D_0=1$ and $D_2=n^2-1$
[note that $D_2$ is the dimension of the adjoint representation of
SU($n$)].  The solution is $D_L=U_L(n/2)$.  For $n>2$, these
dimensions increase exponentially with $L$. Note that these dimensions
are the multiplicities of energy eigenvalues for any of the transfer
matrices/hamiltonians built using the Temperley Lieb algebra (barring
accidental degeneracies).  We thus recover the result (\ref{DL0}) in
the case $\ell=n$. The whole construction (this will be emphasized
more below) is in fact identical when contractible and non
contractible lines carry a different weight, and formally depends only
on the latter; the result can thus be used in the case $\ell \neq n$
as well, reproducing now (\ref{DL0}) in all cases.
 
The commutant algebra can be constructed explicitly as follows. We
introduce the operators (for $k\leq N$)
\be
 \widetilde{J}^{a_1a_2\ldots a_k}_{b_1b_2\ldots b_k}=\sum_{1\leq
 i_1<i_2<\cdots <i_k\leq N} J_{i_1 b_1}^{a_1}J_{i_2
 b_2}^{a_2}\cdots J_{i_k b_k}^{a_k}
\ee
(for $k=0$, we define $\widetilde{J}=1$, and for $k=1$,
$\widetilde{J}^a_b=J^a_b$ as defined earlier). For each $k=0,1,\ldots$
these span a space of dimension $n^{2k}$. In this space of operators
we can impose linear conditions, that the contraction of one of the
indices $a$ with a {\em neighboring} index $b$ (i.e.\ of $a_l$ with
$b_{l\pm 1}$, for $l=1$, $2$, \ldots, $k$) is zero. This gives us a
basis set $J^{a_1\ldots a_k}_{b_1\ldots b_k}$, that are ``traceless'',
in this sense.  For example, for $k=2$, we have
\be
J^{a_1 a_2}_{b_1 b_2}=\widetilde{J}^{a_1 a_2}_{b_1 b_2}%
-\frac{1}{n}\widetilde{J}^{a a_2}_{b_1 a}\delta^{a_1}_{b_2}%
-\frac{1}{n}\widetilde{J}^{a_1b}_{b b_2} \delta^{a_2}_{b_1}%
+\frac{1}{n^2}\widetilde{J}^{a b}_{b
  a}\delta^{a_1}_{b_2}\delta^{a_2}_{b_1}%
\label{commdef}
\ee
and these span a space of dimension $(n^2-1)^2$. In general, the
dimension reads $(D_{k/2})^2$, with $D_{k/2}$ numbers which will turn
out to be the familiar ones. The exact forms are
\be
J^{a_1a_2\ldots a_k}_{b_1b_2\ldots b_k}=(P^\bullet P_\bullet
\widetilde{J})^{a_1a_2\ldots a_k}_{b_1b_2\ldots b_k},
\ee
where $P^\bullet$ ($P_\bullet$) is the (Jones-Wenzl) projection
operator to the ``traceless'' sector on the vector space indexed
by $(a_1,b_2,\ldots)$ [resp., $(b_1,a_2,\ldots,)$], which can be
constructed recursively using the $TL_N(n)$ algebra in these
spaces.

One can show that:
\begin{enumerate}
\item all $J^{a_1a_2\ldots a_k}_{b_1b_2\ldots b_k}$ commute with all
  the $e_i$, hence with all of $TL_{N}(n)$ (they leave the patterns
  unchanged);
\item all $J^{a_1a_2\ldots a_k}_{b_1b_2\ldots b_k}$ with $k>2j$
  annihilate the $j$th representation of the commutant algebra;
\item the space of $J^{a_1a_2\ldots a_k}_{b_1b_2\ldots b_k}$s with
  $k=2j$ acts as the matrix algebra $M_{D_j}({\bf C})$ on the $j$th
  representation; 
\item $J^{a_1a_2\ldots a_k}_{b_1b_2\ldots b_k}$ with
  $k<2j$ map the $j$th representation into itself, and hence in that
  subspace can be written as linear combinations of those with $k=2j$.
  In particular, in our chain of $N$ sites, the operators with $k$ odd
  are linear combinations of those with $k$ even. Hence only even $k$
  are needed.
\end{enumerate}
These results show that {\em the algebra spanned by $J^{a_1a_2\ldots
    a_k}_{b_1b_2\ldots b_k}$ ($k=0,2,\ldots$) is the commutant algebra
  $\cA_n(N)$ of $TL_{N}(n)$ in $V$}, with dimension ${\rm dim\,
}\cA_n(N)=\sum_j (D_j)^2$. Because the dimensions $D_j$ are
independent of $L$, the limit $N\to\infty$ exists, and we write
$\cA_n=\lim_{L\to\infty}\cA_m(N)$. For $m>2$, our algebra $\cA_n$ is
strictly larger than the ``obvious'' global symmetry algebra, which is
the Universal Enveloping Algebra $U({\rm sl}_n)$, which is a proper
subalgebra of $\cA_n$ \cite{RS07-1}. It is not known whether there is
a ``small'' or ``simple'' set of generators for $\cA_n$ (that would be
analogous to the set of $J^a_b$ for $U({\rm sl}_n)$).  $\cA_n$ is not
the Yangian of sl$_n$.

\subsection{One-boundary case}

The one-boundary case can now be interpreted straightforwardly by
constraining the variables on the $i=1$ site to be restricted to a
smaller space ${\bf C}^{n_{\rm l}}$, $n_{\rm l}\leq n$. The most
convenient way to do so is to introduce a {\sl projector} on that
subspace, which obeys exactly the $b_{\rm l}$ relations given in
(\ref{TL1}).

The commutant of the 1BTL algebra is smaller than the algebra
$\cA_{n}(N)$. It is easy to show that it is generated by the same set
of operators $J$ as before, subject to the constraint that the labels
$a_{1},b_{1}$ take values in the restricted subset $a_{1},b_{1}\leq
n_{\rm l}$, {\sl or in the complementary subset} $n_{\rm
  l}<a_{1},b_{1}\leq n$.

The smallest representations of the commutant come from the generators
with four indices, and these occur in numbers, respectively $(n_{\rm
  l}\times n)-1$ and $[(n-n_{\rm l})\times n]-1$. These two numbers
coincide with $D^{b}_2$ and $D^{u}_2$ respectively.

The combinatorics for the higher generators of the symmetry works as 
before: the recurrence relation for the dimensions is not affected by 
the restrictions on the first site since the contractions only involve 
labels on sites $i>1$:
\begin{eqnarray}
    D^{\rm b}_L(n^{2}-1)=D^b_{L+2}+D^b_L+D^b_{L-2}\nonumber\\
    D^{\rm u}_L(n^{2}-1)=D^u_{L+2}+D^u_L+D^u_{L-2}
\end{eqnarray}
and the solution is easily seen to coincide with (\ref{cheby}) in the
case $\ell=n$. The relations extend as well to the more general case,
and one simply has to substitute $\ell$ for  $n$ and $\ell_{\rm l}$ for $n_{\rm l}$.

\subsection{Two-boundary case}

We can now extend this construction to the case of two boundaries, 
meaning we now wish to project the  spin at $i=1$ onto a subspace of 
dimension $n_{\rm l}$ and the  one at $i=N$ onto a subspace of 
dimension $n_{\rm r}$. An important distinction will be whether these two 
subspaces have an empty intersection or not. 

If the intersection is empty, the generators $J$ with two indices can
be built with no trace subtraction necessary when the first labels
belong to the subset of $n_{\rm l}$ colors and the second one to the
subset with $n_{\rm r}$ colors. This gives a first representation of
the commutant, of dimension $D^{\rm bb}_2=n_{\rm l}n_{\rm r}$. If the
first labels belong to the orthogonal subset with $n-n_{\rm l}$
colors, there will necessarilty be overlap with the $n_{\rm r}$ colors
from the other subset, hence a trace subtraction is necessary, whence
$D^{\rm ub}_2=(n-n_{\rm l})n_{\rm r}-1$, and $D^{\rm bu}_2=n_{\rm l}
(n-n_{\rm r})-1$.
Meanwhile for the case where the two labels belong to the complements
with $n-n_{\rm l}$ and $n-n_{\rm r}$ colors respectively, we expect as
well to have an overlap, and thus $D^{\rm uu}_2=(n-n_{\rm l})(n-n_{\rm
  r})-1$.

Fusion cannot be carried out naively any longer, since now both sides
of the chain are affected. We can however imagine performing fusion of
two representations of the commutant in the one-boundary case, one of
them corresponding to having the leftmost index restricted and the
other one the rightmost index restricted. This should give rise to
representations of the commutant where now both left and right indices
are restricted. For instance one can consider
\begin{equation}
    \left.D^{\rm b}_2\right|_{n_{\rm l}}\times\left.  D^{\rm b}_L\right|_{n_{\rm r}}
\label{somelhs}
\end{equation}
where the second term can be interpreted as coming from the system
where the labels on the rightmost site are restricted (though this
does not make any difference algebraically). Performing the
contractions, we see that we can get generators with $L+2$ or $L$
indices (corresponding pictorially to contracting zero or one pair of
lines) with one of the indices restricted on the left remaning. We can
also contract two pairs of indices, killing the restricted index on
the left. This translates into the right-hand side of (\ref{somelhs}) being
\begin{equation}
	\left.D^{\rm bb}_{L+2}\right|_{n_{\rm l},n_{\rm r}}+
	\left.D^{\rm bb}_L\right|_{n_{\rm l},n_{\rm r}}+\left.D^{\rm b}_{L-2}\right|_{n_{\rm r}}
\end{equation}
recovering the relation (\ref{fusionidentity}) found in the previous section. 

Finally, one could also decide (as in section \ref{doubly_blobbed})
that the $n_{\rm l}$ and $n_{\rm r}$ colors are not orthogonal but
have some overlap.  Then a loop touching both the left and right
boundaries gets the weight $n_{\rm b}$. Other formulas follow easily.
For instance, the dimension $D^{\rm bb}_2=n_{\rm l}n_{\rm r}-1$ now.
Meanwhile, if $n_{\rm l}$ and $n_{\rm r}$ have some overlap, their
complements also will generically, and so will the complement of one
with the other.  Therefore we expect the other dimensions to remain
unchanged.

It is convenient to give uniform parametrizations for all the
dimensions. Set
\begin{eqnarray}
    \ell&=&2\cosh\alpha\nonumber\\
    \ell_{\rm l}&=&{\sinh(\alpha+\beta_{\rm l})\over \sinh(\beta_{\rm l})}=D_1^{\rm b}\nonumber\\
    \ell-\ell_{\rm l}&=&{\sinh(\beta_{\rm l}-\alpha)\over\sinh\beta_{\rm l}}=D_1^{\rm u}
\end{eqnarray}
Then we have
\begin{equation}
D_L={\sinh(L+1)\alpha\over \sinh\alpha}
\end{equation}
together with
\begin{eqnarray}
D^{\rm b}_{L}={\sinh(L\alpha+\beta_{\rm l})\over\sinh\beta_{\rm l}}\nonumber\\
D^{\rm u}_{L}={\sinh(L\alpha-\beta_{\rm l})\over\sinh(-\beta_{\rm l})}
\end{eqnarray}
and 
\begin{eqnarray}
D^{\rm bb}_L&=&{\sinh[(L-1)\alpha+\beta_{\rm l}+\beta_{\rm r}]\sinh\alpha\over\sinh \beta_{\rm l}\sinh \beta_{\rm r}}\nonumber\\
D^{\rm uu}_L&=&{\sinh[(L-1)\alpha-\beta_{\rm l}-\beta_{\rm r}]\sinh\alpha\over\sinh \beta_{\rm l}\sinh \beta_{\rm r}}\nonumber\\
D^{\rm ub}_L&=&-{\sinh[(L-1)\alpha-\beta_{\rm l}+\beta_{\rm r}]\sinh\alpha\over\sinh\beta_{\rm l}\sinh\beta_{\rm r}}\nonumber\\
D^{\rm bu}_L&=&-{\sinh[(L-1)\alpha+\beta_{\rm l}-\beta_{\rm r}]\sinh\alpha\over\sin \beta_{\rm l}\sinh \beta_{\rm r}}
\end{eqnarray}
where we have introduced the parameters $\beta_{\rm r}$ in analogy with
$\beta_{\rm l}$ in an obvious fashion. Notice that one goes from the
sector b to the sector u by changing the sign of the parameters
$\beta_{\rm l},\beta_{\rm r}$. For instance one checks that $D_3^{\rm
  bb}=\ell\ell_{\rm l}\ell_{\rm r}-\ell_{\rm l}-\ell_{\rm r}$; $D_3^{\rm
  uu}=\ell^3-2\ell-(\ell_{\rm l}+\ell_{\rm r})(\ell^2-1)+\ell\ell_{\rm l}\ell_{\rm r}$;
$D_1^{\rm uu}=\ell-\ell_{\rm l}-\ell_{\rm r}$.

\section{Determinants}

One major motivation for deriving the dimensions $D_L$ is to find out when they vanish. Recalling Eqs.~(\ref{defK0})--(\ref{ZDK}) we see that
when $D_L=0$ for some $L$ there is a whole subset of eigenvalues which does no
longer contribute to the partition function.

The effects of this in the 0BTL case are well understood. Note that
(\ref{DL0}) can be rewritten
\be
 D_L = U_L(\ell/2) = \prod_{j=1}^{L/2} \left(\ell^2 - B^{(j)}_{L+1} \right)
\ee
where $B^{(j)}_k = (q+q^{-1})^2$ with $q={\rm e}^{i \pi j/k}$. Thus,
we may have $D_L=0$ provided that $\ell=\pm(q+q^{-1})$, with the usual
quantum group deformation parameter $q$ being a rational-order root of
unity.

\begin{figure}
  \begin{center}
    \includegraphics[scale=0.4]{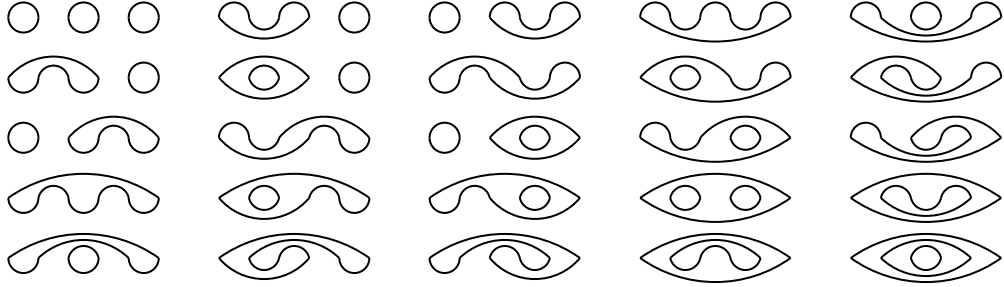}
  \end{center}
  \caption{Gram matrix for the 0BTL case with $N=6$ and $L=0$.}
  \label{fig:meander}
\end{figure}

Another way to detect such singular behavior is through the study of
the Gram matrix ${\cal G}^{\rm M}_{N,L}$ whose rows and columns are indexed
by the reduced states of section~\ref{sec:states0} (see
Fig.~\ref{fig:red_states0}), and whose entries are the values of the
corresponding inner product. For instance, with $N=6$ strands and
$L=0$ strings we get the matrix written pictorially in
Fig.~\ref{fig:meander} and algebraically as follows:
\be
 {\cal G}^{\rm M}_{6,0} = \left[ \begin{array}{ccccc}
 n^3 & n^2 & n^2 & n   & n^2 \\
 n^2 & n^3 & n   & n^2 & n   \\
 n^2 & n   & n^3 & n^2 & n   \\
 n   & n^2 & n^2 & n^3 & n^2 \\
 n^2 & n   & n   & n^2 & n^3 \\
 \end{array} \right]
\ee
Its determinant turns out to be a product of powers of the amplitudes (\ref{DL0}),
$\det {\cal G}^{\rm M}_{6,0} = D_1^4 D_2^4 D_3$, where we have set $\ell=n$.
The formula for arbitrary $N$ has been proven \cite{meander} to be
\be
 \det {\cal G}^{\rm M}_{N,0} = \prod_{m=1}^{N/2} (D_m)^{{ N \choose N/2-m} -
  2 {N \choose N/2-m-1} + {N \choose N/2-m-2}}
\ee
The superscript ${\rm M}$ indicates that this determinant is
encountered in the theory of meanders.

So the general idea is that zeroes of the Gram determinants correspond
to the appearance of invariant subspaces, and thus should also
correspond to states disappearing in the partition functions, i.e., to
zeroes of the degeneracies $D$. It is then natural to expect that the
determinants \cite{jamesmurphy} should factor out nicely in terms of the $D$ numbers,
also for the 1BTL and 2BTL cases. Assuming this and calculating the
determinants formally and numerically for small enough sizes
(typically for $N \le 12$) allowed us to obtain a number of
conjectures for all values of $N$.

\subsection{Conventions}
\label{sec:conventions}

Before stating our main results, we should point out that there is not
just one but several ways of defining the Gram matrices of interest.

It is natural to consider matrices ${\cal G}_{N,L}$ on $N$ strands
using only reduced states with a fixed number $L$ of strings. Consider then
first the case of $L=0$. There are then two choices to be made:
\begin{enumerate}
\item One may work with the full algebra [i.e., with $\lambda_{\rm l}
  \neq 0$ in the 1BTL case (\ref{TM1}), or $\lambda_{\rm l},
  \lambda_{\rm r} \neq 0$ in the 2BTL case (\ref{TM2})], or constrain
  to the smaller algebra where left/rightmost objects are always
  blobbed [i.e., with $\lambda_{\rm l} = 0$ in the 1BTL case, or
  $\lambda_{\rm l} = \lambda_{\rm r} = 0$ in the 2BTL case].
\item One may use the reduced states defined in the present paper, or
  the alternate states defined in \cite{MS} (with each link projected
  on $b_{\rm l}$ or on $1-b_{\rm l}$) that we have briefly discussed in section
  \ref{sec:states1}.
\end{enumerate}
It is easy to see that choice 2 does not change the result for the
determinant, as it is simply a change of basis. However, the determinant clearly depends on choice 1,
although the two possibilities are closely related. We shall illustrate
this below in the 1BTL case.

\begin{figure}
  \begin{center}
    \includegraphics[scale=0.4]{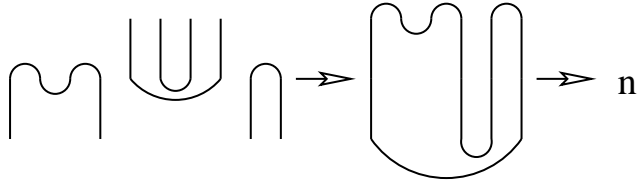}
  \end{center}
  \caption{``Semimeander'' convention for inner products in the
    presence of strings.}
  \label{fig:semimeander}
\end{figure}

When strings are present ($L \neq 0$) the situation is more involved.
Consider first for simplicity the 0BTL case. There are then more
choices to be made:
\begin{enumerate}
\addtocounter{enumi}{2}
\item One may require that the inner product is zero if the strings in
  the top and bottom reduced states are not in the same positions. We
  shall however not impose this requirement in the following, since it
  will roughly speaking divide the arc system into several
  non-interacting parts, and hence lead to a rather trivial
  factorization of the determinant.
\item One may require or not the conservation of lines. By line conservation
  we understand that each string on the bottom connects (eventually via some
  intermediary arcs) to a string on the top.
\item When there is no line conservation, one may chose a non-trivial rule
  for associating Boltzmann weights to the strings.
\end{enumerate}
Clearly, imposing line conservation in point 4 is akin to the
construction of transfer matrix sectors with a conserved number of
strings (see section \ref{sec:states0}). In point 5, an example of a
non-trivial rule is to connect the $k$'th leftmost top string to the
$k$'th leftmost bottom string, and weighting by a factor of $n$ for
each of the loops thus formed. This is illustrated in
Fig.~\ref{fig:semimeander}. In Ref.~\cite{semimeander}, the
corresponding Gram determinant was referred to as the ``semimeander
determinant'' and proven to be
\begin{eqnarray}
 \det {\cal G}^{\rm SM}_{N,L} &=& \prod_{m=1}^{\frac{N-L}{2}+1} (D_m)^{c_{N,2m+L}-c_{N,2m+2+L}+L(c_{N,2m+L-2}-c_{N,2m+L})} \label{semidet} \\
 c_{k,h} &=& {k \choose \frac{k-h}{2}} - {k \choose \frac{k-h}{2}-1} \nonumber
\end{eqnarray}

We have found formulae for the 1BTL and 2BTL determinants with a
variety of choices 3--5. It does however not seem urgent to make all
of these formulae appear on print. We therefore adopt in what follows
the choices that we have found lead to the simplest results for the
boundary determinants, viz.:
\begin{enumerate}
\addtocounter{enumi}{2}
\item Top and bottom strings are {\em not} required to be in the same
  positions.
\item Line conservation is imposed.
\item String carry a trivial Boltzmann weight of one.
\end{enumerate}
These choices are consistent with those made in \cite{MS}. As an
example of how they affect the results for the determinants, the 0BTL
result (\ref{semidet}) now becomes instead (the proof of this result
appears in \cite{Paulunpublished})
\be
 \det {\cal G}_{N,L} = \prod_{m=0}^{\frac{N-L}{2}-1}
             \left( \frac{D_{\frac{N+L}{2}-m}}{D_{\frac{N-L}{2}-1-m}}
             \right)^{{N \choose m} - {N \choose m-1}}
 \label{semidet1}
\ee
Note that despite of the way we have written (\ref{semidet1}) there are
of course no poles in the expression.

\begin{figure}
  \begin{center}
    \includegraphics[scale=0.4]{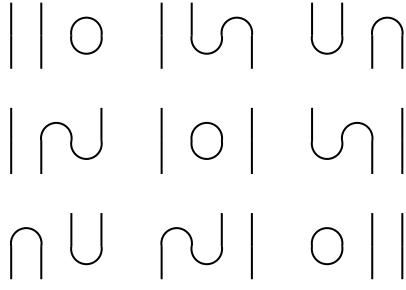}
  \end{center}
  \caption{Gram matrix for the 0BTL case with $N=4$ and $L=2$.}
  \label{fig:semidetex}
\end{figure}

It is instructive to compare (\ref{semidet})--(\ref{semidet1}) through
the explicit example $N=4$ and $L=2$. The Gram matrix is written
pictorially in Fig.~\ref{fig:semidetex}. Algebraically it reads, with the
semimeander convention of (\ref{semidet}),
\begin{equation}
 {\cal G}^{\rm SM}_{4,2} = \left[ \begin{array}{ccc}
 n^3 & n^2 & n   \\
 n^2 & n^3 & n^2 \\
 n   & n^2 & n^3 \\
 \end{array} \right]
\end{equation}
and has determinant $D_1^5 D_2^2 = n^5(n^2-1)^2$. With the conventions
of (\ref{semidet1}), Fig.~\ref{fig:semidetex} reads algebraically
\begin{equation}
 {\cal G}_{4,2} = \left[ \begin{array}{ccc}
 n & 1 & 0   \\
 1 & n & 1 \\
 0 & 1 & n \\
 \end{array} \right]
\end{equation}
and has determinant $D_3 = n^3-2n$.

Finally, for the cases with boundaries, we define as zero any inner
product that does not satisfy the sector structure (${\rm u} =$
unblobbed, ${\rm b} =$ blobbed) that we have developped for the
transfer matrix. This means that the same sector is used for the
reduced states on top and bottom. Moreover, an unblobbed string is
not allowed to connect onto a blobbed arc. As a matter of notation,
the sector labels are shown as superscripts on the Gram matrix, i.e.,
${\cal G}^\alpha_{N,L}$ for the 1BTL case, and ${\cal G}^{\alpha
  \beta}_{N,L}$ for the 2BTL case.

\subsection{One-boundary case}
  
The determinants of the Gram matrix for the 1BTL algebra were
determined in \cite{MS}. There, the whole algebra was
considered, corresponding to $\lambda_{\rm l}\neq 0$ in our notations.
This is the case we will discuss for a while. In the $L=0$ sector
there is no difference between blobbed and unblobbed sectors, and one
has
\begin{equation}
 \det {\cal G}^{\rm b}_{N,0} =  \det {\cal G}^{\rm u}_{N,0} = 
 \prod_{m=1}^{N/ 2} \left(D_m^{\rm u}D_m^{\rm b}\right)^{N\choose N/2-m}
 \label{det1BTLnolines}
\end{equation}
The order of this determinant in terms of the variables $n_{\rm l}$ and $n$
is
$$\sum_{m=1}^{N/ 2} 2m{N\choose N/2-m}={N\over 2}{N\choose N/2}.$$
The total number of diagrams in the sector is ${N\choose N/2}$, and
the matrix elements on the diagonal of the Gram matrix are each of order
${N/2}$, in agreement with this result.

\begin{figure}
  \begin{center}
    \includegraphics[scale=0.4]{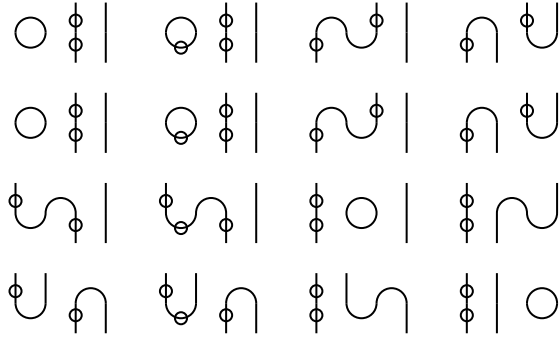}
  \end{center}
  \caption{Gram matrix for the 1BTL case with $N=4$ and $L=2$, blobbed sector.}
  \label{fig:det1BTL}
\end{figure}

Our conventions for the definition of the Gram matrix with $L \neq 0$
have been given in section~\ref{sec:conventions}. Before addressing
the general case, let us illustrate them on the example $N=4$, $L=2$,
blobbed sector. There are four basis states, and the matrix ${\cal
  G}^{\rm b}_{4,2}$ is shown pictorially in Fig.~\ref{fig:det1BTL}.
Algebraically it reads:
\be
 {\cal G}^{\rm b}_{4,2} = \left[ \begin{array}{cccc}
 n & n_{\rm l} & 1 & 0 \\
 n_{\rm l} & n_{\rm l} & 1 & 0 \\
 1 & 1 & n & 1 \\
 0 & 0 & 1 & n \\
 \end{array} \right]
\ee
and one finds $\det {\cal G}^{\rm b}_{4,2} = D^{\rm u}_1 D^{\rm b}_3$,
which indeed factorizes nicely in terms of the amplitudes (\ref{cheby}).

The general result for the blobbed sector is
\begin{equation}
  \det {\cal G}^{\rm b}_{N,L} = \prod_{m=1+L/2}^{N/ 2} \left(D_{m+ L/2}^{\rm b}D_{m-L/2}^{\rm u}\right)^{N\choose {N/2}-m}
\end{equation}
Note that upon setting $L=0$ in this formula one recovers
(\ref{det1BTLnolines}), as a consequence of our choice of conventions.
The order of the determinant is now
$$
\sum_{m=1+L/2}^{N/ 2} 2m {N\choose N/2-m}={N-L\over 2}{N\choose \frac{N-L}{2}}
$$
Notice also the special case $L=N-2$
\begin{equation}
 \det {\cal G}^{\rm b}_{N,N-2} = D_{N-1}^{\rm b}D_1^{\rm u}
\end{equation}
which was discussed in great details  in \cite{MS}.

In the unblobbed sector, we have a similar expression where $D^{\rm u}$ and
$D^{\rm b}$ have been switched:
\begin{equation}
  \det {\cal G}^{\rm u}_{N,L} =
  \prod_{m=1+L/2}^{N/ 2} \left(D_{m-L/2}^{\rm b}D_{m+L/2}^{\rm u}\right)^{{N\choose N/2-m}}
\end{equation}

It is easy to obtain similar expressions for the determinants in the
case of the restricted algebra (i.e., $\lambda_{\rm l}=0$). We find
the following conjectures
\begin{eqnarray}
  \det {\cal G}^{\rm b}_{N,L} &=&
  \prod_{m=1+L/2}^{N/ 2} \left(D_{m+ L/2}^{\rm b}\right)
  ^{N-1\choose N/2-m}\prod_{m=2+L/2}^{N/ 2}  \left(D_{m-1-L/2}^{\rm u}\right)^{N-1\choose {N/2}-m} \nonumber \\
  \det {\cal G}^{\rm u}_{N,L} &=&
  \prod_{m=1+L/2}^{N/ 2} \left(D_{m- L/2}^{\rm b}\right)
  ^{N-1\choose N/2-m}\prod_{m=2+L/2}^{N/ 2}  \left(D_{m-1+L/2}^{\rm u}\right)^{N-1\choose {N/2}-m}
\end{eqnarray}

\subsection{Two-boundary case}

In the 2BTL case we have concentrated on the full algebra (see also 
\cite{towers}), with
$\lambda_{\rm l},\lambda_{\rm r}\neq 0$. Consider first the situation
with $L=0$ strings, where loops are allowed to touch both boundaries
with weight $n_{\rm b}$. In this case all sectors coincide, but it is
notationally convenient to use the sector label uu to recall that there
are two boundaries. We conjecture that
\begin{eqnarray}
  \det {\cal G}^{\rm uu}_{N,0} =
  \prod_{k=1}^{N\over 2}\left[
  \left(n_{\rm b}+\sum_{m=1}^k D^{\rm uu}_{2m-1}\right)
  \left(n_{\rm b}+\sum_{m=1}^k D_{2m-1}^{\rm bb}\right)
  \right.\nonumber\\
  \left.
  \left(n_{\rm b}-\sum_{m=1}^k D_{2m-1}^{\rm ub}\right)
  \left(n_{\rm b}-\sum_{m=1}^k D^{\rm bu}_{2m-1}\right)\right]^{a_{N,k}}
  \label{det2BTLnb}
\end{eqnarray}
where we have defined
\begin{equation}
  a_{N,k}=\sum_{m=0}^{N/2-k} {N \choose m}
\end{equation}
As this work was nearing completion, we became aware of the recent
paper by de Gier and Nichols where a very similar formula for $\det
{\cal G}^{\rm uu}_{N,0}$ is derived (see Theorem 5.3 of \cite{GN}).
This latter result was established in the word basis rather than the
diagram one, and as a result differs from (\ref{det2BTLnb}) by a simple
prefactor.

Note that the sums of amplitudes appearing in (\ref{det2BTLnb}) can be
rewritten in terms of Chebyshev polynomials as follows [we suppress the
argument $n/2$ in $U_L(n/2)$]:
\begin{eqnarray}
 \sum_{m=1}^k D^{\rm uu}_{2m-1} &=&
 U_k U_{k-1} - (n_{\rm l}+n_{\rm r})(1+U_k U_{k-2}) +
 n_{\rm l} n_{\rm r} U_{k-1} U_{k-2} \nonumber \\
 \sum_{m=1}^k D^{\rm bb}_{2m-1} &=&
 n_{\rm l} n_{\rm r} U_k U_{k-1} - (n_{\rm l}+n_{\rm r})(1+U_k U_{k-2}) +
 U_{k-1} U_{k-2} \nonumber \\
 \sum_{m=1}^k D^{\rm ub}_{2m-1} &=&
 n_{\rm r}(1+U_k U_{k-2}) - (1+n_{\rm l} n_{\rm r})U_{k-1}U_{k-2} +
 n_{\rm l}(1+U_{k-1}U_{k-3}) \nonumber \\
 \sum_{m=1}^k D^{\rm bu}_{2m-1} &=&
 n_{\rm l}(1+U_k U_{k-2}) - (1+n_{\rm l} n_{\rm r})U_{k-1}U_{k-2} +
 n_{\rm r}(1+U_{k-1}U_{k-3})
 \label{listmagic}
\end{eqnarray}

Consider next the situation with $L \neq 0$, so that no loop can
touch both boundaries. We find then the following conjectures
\begin{eqnarray}
 \det {\cal G}^{\rm uu}_{N,L} &=&
 \prod_{m=0}^{{N-L\over 2}-1}
 \left(D^{\rm uu}_{{N+L\over 2}-m} D^{\rm b(l)}_{{N-L\over 2}-m}
 D^{\rm b(r)}_{{N-L\over 2}-m} D_{{N-L\over 2}-m-1}\right)^{a_{N,N/2-m}}
 \nonumber \\
 \det {\cal G}^{\rm ub}_{N,L} &=&
 \prod_{m=0}^{{N-L\over 2}-1}
 \left(D^{\rm ub}_{{N+L\over 2}-m} D^{\rm b(l)}_{{N-L\over 2}-m}
 D^{\rm u(r)}_{{N-L\over 2}-m} D_{{N-L\over 2}-m-1}\right)^{a_{N,N/2-m}}
 \nonumber \\
 \det {\cal G}^{\rm bu}_{N,L} &=&
 \prod_{m=0}^{{N-L\over 2}-1}
 \left(D^{\rm bu}_{{N+L\over 2}-m} D^{\rm u(l)}_{{N-L\over 2}-m}
 D^{\rm b(r)}_{{N-L\over 2}-m} D_{{N-L\over 2}-m-1}\right)^{a_{N,N/2-m}}
 \nonumber \\
 \det {\cal G}^{\rm bb}_{N,L} &=&
 \prod_{m=0}^{{N-L\over 2}-1}
 \left(D^{\rm bb}_{{N+L\over 2}-m} D^{\rm u(l)}_{{N-L\over 2}-m}
 D^{\rm u(r)}_{{N-L\over 2}-m} D_{{N-L\over 2}-m-1}\right)^{a_{N,N/2-m}}
\label{det2BTLstrings}
\end{eqnarray}
We have verified that the result (\ref{det2BTLstrings}) for $\det
{\cal G}^{\rm uu}_{N,L}$ holds true also for $L=0$, provided that we
explicitly forbid the loops to touch both boundaries.%
\footnote{Note the consistency with the label uu assigned to the other
  $L=0$ formula (\ref{det2BTLnb}).}
By this we mean that not only do we omit doubly blobbed arcs in the
basis of reduced states, but we also set to zero any inner product
that involves adjoining a left-blobbed and a right-blobbed arc.

The results (\ref{det2BTLstrings}) do not appear in \cite{GN}, but can
presumably be proven by similar methods.

The interpretation of the determinant (\ref{det2BTLnb}) in the $L=0$
sector is not obvious, except for the first term which reads $n_{\rm
  b}(n-n_{\rm l}-n_{\rm r}+n_{\rm b})(n_{\rm l}-n_{\rm b})(n_{\rm
  r}-n_{\rm b})$.
Indeed, if we interpret $n_{\rm b}$ as the number of colors common to
both boundary conditions, a loop touching both boundaries can have
$n_{\rm b}$ colors, a loop touching only the left boundary is allowed
$n_{\rm l}-n_{\rm b}$ colors (since among the $n_{\rm l}$ ones,
$n_{\rm b}$ are still allowed on the right side), a loop forbidden to
touch either boundary has $n-n_{\rm l}-n_{\rm r}+n_{\rm b}$ colors
since by subtracting $n_{\rm l}$ and $n_{\rm r}$ we subtracted the
$n_{\rm b}$ common ones twice. No such interpretation seems possible
for the other terms.  A careful study of the expression
(\ref{det2BTLnb}) shows that it is useful to parametrize $n_{\rm b}$
as
\begin{equation}
  n_{\rm b}={\sinh{\beta_{\rm l}+\beta_{\rm r}+\alpha+\beta_{\rm b}\over 2}
    \sinh{\beta_{\rm l}+\beta_{\rm r}+\alpha-\beta_{\rm b}\over 2}\over \sinh\beta_{\rm l}\sinh \beta_{\rm r}}
\end{equation}
The determinant can then be rewritten as
\begin{equation}
  \det {\cal G}^{\rm uu}_{N,0} = \prod_{k=1}^{N/2} \prod_{\epsilon_{\rm l,r,b}=\pm 1}\left[{\sinh {1\over 2}[(2k-1)\alpha+\epsilon_{\rm l}\beta_{\rm l}+
      \epsilon_{\rm r}\beta_{\rm r}+\epsilon_{\rm b}\beta_{\rm b}]
      \sinh\alpha\over \sinh \beta_{\rm l}\sinh\beta_{\rm r}}\right]^{a_{N,k}}
\end{equation}
This form exhibits the hidden
symmetry between $\beta_{\rm l},\beta_{\rm r}$ and $\beta_{\rm b}$.

The combinations appearing in the determinant suggest that it is
natural to take yet other quotients of the algebra, making it smaller
and thus the commutant bigger. A combination such as
$$
n_{\rm b}+\sum_{m=1}^k D^{\rm bb}_{2m-1}
$$
is naturally associated to an algebraic object where contractions of
the left and right most strings with their immediate neighbours would
be allowed. Why this would have to be so deserves further study.

Another elusive feature of Eq.~(\ref{det2BTLnb}) is that the ``magic''
values of $n_{\rm b}$ (i.e., that cause the determinant to vanish) are not
revealed by the vanishing of the eigenvalue amplitudes $D$ (see
section~\ref{doubly_blobbed}).

Let us remark that Nichols \cite{N} has studied the
issue of magic values of $n_{\rm b}$ for the cases where the 2BTL
model reduces to the Potts model with $Q=2$ and $Q=3$ states with
various boundary conditions expressed in terms of the Potts spins.%
\footnote{To convert the notation of \cite{N} into the one used
  here, note that his $s_0$ is our $b_{\rm l}/n_{\rm l}$, his $s_N$ is our
  $b_{\rm r}/n_{\rm r}$, and his $b$ is our $b/(n_{\rm l} n_{\rm r})$.}
Since these are minimal models, it is not surprising that the list
(\ref{listmagic}) of magic $n_{\rm b}$ is finite in these cases. Our
list coincides with that of \cite{N}, except that we are of
course unable to distinguish boundary conditions which differ by a
permutation of spin labels.

\section{Conclusion}

In conclusion, we expect that these results should have a natural interpretation in the (boundary) conformal field theory framework, and we hope to discuss this soon.
\bigskip

\section*{Acknowledgments}

This work was supported through the European Community Network ENRAGE
(grant MRTN-CT-2004-005616) and by the Agence Nationale de la
Recherche (grant ANR-06-BLAN-0124-03). Part of this work was done
while the authors participated the Random Shapes programme at the
Institute for Pure and Applied Mathematics (IPAM/UCLA).

\end{document}